\documentclass[fleqn,usenatbib]{mnras}

\usepackage{newtxtext}
\usepackage{mathptmx}

\usepackage[T1]{fontenc}

\DeclareRobustCommand{\VAN}[3]{#2}
\let\VANthebibliography\thebibliography
\def\thebibliography{\DeclareRobustCommand{\VAN}[3]{##3}\VANthebibliography}

\usepackage{hyperref}
\usepackage{academicons}
\usepackage{booktabs}
\usepackage[dvipsnames,table,xcdraw]{xcolor}

\usepackage{graphicx}	
\usepackage{subcaption} 
\usepackage{amsmath}	
\usepackage{orcidlink}

\usepackage{float}

\newcommand{\orcid}[1]{\href{https://orcid.org/#1}{\textcolor[HTML]{A6CE39}{\aiOrcid}}}

\title[PUCHEROS+]{The evolution of PUCHEROS from a basic to a competitive tool for stellar astrophysics}

\author[L. Antonucci et al.]{
L. Antonucci\thanks{E-mail: leantonucci@uc.cl}\orcidlink{0009-0003-4135-1296}$^{1,2}$
L. Vanzi$^{1,2}$,
A. A. Zapata\orcidlink{0000-0003-2326-6488}$^{1,2}$,
M. Flores$^{1,2}$,
A. Suárez$^{1,2}$,
R. Brahm$^{3}$,\newauthor
T. Shen$^{1,4}$,
M. Parra$^{1,4}$,
R. Ormazábal$^{1,2}$,
G. Avila$^{5}$,
P. Kab\'{a}th$^{6}$,
A. Hatzes$^{7}$,
P. Gajdo\v{s}\orcidlink{0000-0003-1478-3256}$^{6,8}$,\newauthor
M. Skarka\orcidlink{0000-0002-7602-0046}$^{6}$,
J. \v{Z}\'{a}k$^{6}$,
P. Odert$^{9}$,
J. Lipt\'{a}k$^{6,10}$,
R. Greimel$^{11}$,
and M. Leitzinger$^{9}$
\\
$^{1}$Center of Astro Engineering, Pontificia Universidad Católica de Chile, Av. Vicuña Mackenna 4860, 782-043 Santiago, Chile\\
$^{2}$Department of Electrical Engineering, Pontificia Universidad Católica de Chile, Av. Vicuña Mackenna 4860, 782-043 Santiago, Chile\\
$^{3}$Facultad de Ingeniería y Ciencias, Universidad Adolfo Ibañez, Av. Diag. Las Torres 2640, Peñalolén, 7941169 Santiago, Chile\\
$^{4}$ BlueShadows, Av. Alfredo Barros Errazuriz 1954 Of. 706, 7500521 Santiago, Chile \\
$^{5}$ ESO-European Organisation for Astronomical Research in the Southern Hemisphere, Casilla 19001, Santiago 19, Chile\\
$^{6}$Astronomical Institute, Czech Academy of Sciences, Fri\v{c}ova 298, 251 65 Ond\v{r}ejov, Czechia\\
$^{7}$Thüringer Landessternwarte Tautenburg, D-07778 Tautenburg, Germany\\
$^{8}$Institute of Physics, Faculty of Science, Pavol Jozef \v{S}af\'arik University, Park Angelinum 9, 040 01 Ko\v{s}ice, Slovakia\\
$^{9}$ Institute of Physics, Department for Astrophysics and Geophysics, University of Graz, Universit\"{a}tsplatz 5, A-8010 Graz, Austria\\
$^{10}$ Astronomical Institute of Charles University, V Hole\v{s}ovi\v{c}k\'{a}ch 2, CZ-180 00 Prague, Czech Republic\\
$^{11}$ RG Science, Schanzelgasse 17, A-8010 Graz, Austria
}

\date{Accepted XXX. Received YYY; in original form ZZZ}

\pubyear{2024}

\begin{document}
\label{firstpage}
\pagerange{\pageref{firstpage}--\pageref{lastpage}}
\maketitle

\begin{abstract}We present PUCHEROS+, a new spectrograph developed as an enhanced version of PUCHEROS, which was the first high-resolution spectrograph
built at the Pontificia Universidad Catolica de Chile (UC). With respect to its predecessor, PUCHEROS+ includes a substantial number of
improvements, mainly: a new scientific detector, improved objective optics, calibration system, guiding, active thermal control and remote observing mode. These upgrades convert our early prototype into a much more powerful instrument for science. 
With a spectral resolution of $R=18{,}000$, a spectral range between $400$ nm and $730$ nm and an instrument efficiency of about $30$\%, PUCHEROS+ was tested at the ESO 1.52\,m telescope where it has reached a limiting magnitude of about $12$ in V-Band and radial velocity (RV) precision of about $30$ m\,s$^{-1}$. The instrument was conceived as a pathfinder for the high-resolution echelle spectrograph PLATOSpec. At the same time, it demonstrates that a compact, relatively low-cost spectrograph can be efficiently employed for long term monitoring campaigns and as support facility for space missions, in particular if operated remotely at relatively small or medium size telescopes.
\end{abstract}

\begin{keywords}
instrumentation: spectrographs–techniques: radial velocities–techniques: spectroscopic–planetary systems.
\end{keywords}



\section{Introduction}\label{section: sec1 introduction}

PUCHEROS (Pontificia Universidad Cat\'olica High Echelle Resolution Optical Spectrograph)
was developed at the AIUC (Center of Astro Engineering of the Pontificia Universidad Cat\'olica) with the objective of providing high-resolution spectroscopy for small to medium size
telescopes at an affordable price \citep{Vanzi2012}. The instrument was conceived as an aid for teaching and for carrying out research, mostly on bright stellar targets. The reduced budget, at the time, forced a number of strong
limitations in design and performance. However, with additional resources and experience we can now present a new design where most of the early limitations have been overcome. This  produces an enhanced performance with much improved science capabilities. In addition, our compact, relatively low cost but high performance spectrograph is optimized for remote observing. Currently, there is a special need for monitoring facilities which can acquire long time series for space missions such as \textit{TESS} \citep{Ricker2014} and PLATO \citep{rauer2016plato}.

This paper is organized as follows: in section \ref{section: sec2 PUCHEROS} we present a brief description of the original PUCHEROS spectrograph and the limitations which motivated the current investment into a new improved version. Section \ref{section: sec3 PUCHEROS+} describes PUCHEROS+ with its new characteristics and its associated systems. In section \ref{section: sec4 Results} we present the performance of the instrument at the ESO 1.52\,m telescope. Section \ref{section: science} showcases five cases of scientific demonstration. In section \ref{section: sec5 Discussion} we discuss the results and finally in section \ref{section: sec6 Conclusions} we present conclusions and future perspectives.
\section{PUCHEROS}\label{section: sec2 PUCHEROS}

PUCHEROS is a high-resolution fiber-fed echelle spectrograph. It was conceived prioritizing simplicity and minimizing cost, finding the best compromise of performance and efficiency. It is primarily used as an academic tool to be employed at the university observatory. Most components were chosen to be commercially available off-the-shelf. The collimator is a parabolic mirror from Edmund Optics with a focal length of $647.7$\,mm. The diameter of the collimated beam is $33$\,mm. The echelle grating has $44.41$\,gr\,mm$^{-1}$, a blaze angle of $70^\circ$ and a size of $100\times50$\,mm, manufactured by Richardson. The cross dispersion is provided by two prisms of $SF2$ and aperture angle of $48^\circ$, the only components custom designed and made. The objective has an equivalent focal length of $221$\,mm and it is obtained by combining two commercial lenses, an achromatic doublet and a meniscus lens as field corrector. The detector is a Finger Lake Instrumentation FLI ProLine PL1001E, which is equipped with a $1K \times 1K$ CCD with pixels of $24\,\mu$m, front illuminated, based on the Kodak KAF-1001E sensor. The instrument covers the visible spectral range ($390-730$\,nm) with a spectral resolution of approximately R=$18{,}000$. The fiber that collects light from the telescope and injects it into the spectrograph has a core diameter of $50\,\mu$m. PUCHEROS was installed at the $50$\,cm telescope of the UC Observatory Santa Martina (formerly ESO50 at La Silla) at the beginning of 2011 and has been in operation since.\\

The main limitations of PUCHEROS in order of decreasing severeness are:
\begin{enumerate}
\item \textbf{Detector.} The choice of a cheap detector implies several strong limitations. A front
illuminated CCD has less than optimal QE, with a peak of nearly $70$\% at $550$\,nm, which drops below $40$\% at $400$\,nm. The cooling power is limited and the detector is being
operated at $-35^{\circ}C$, with an A/D rate of $1$\,MHz. The cooling is provided by a Peltier stage and the heat extraction is obtained through air flow. This, together with the use of a standard monostable shutter, considerably limits the possibility of stabilizing the temperature of the instrument. As a consequence of the operation temperature, dark signal ($0.02$\,e$^-$\,pix$^{-1}$\,s$^{-1}$) and read-out noise ($10$\,e$^-$) are higher than the typical values of devices
operated at lower temperatures. However, possibly the most severe limitation is a strong memory effect produced by Residual Bulk Image (RBI) \citep{crisp2009residual}. This makes necessary careful planning of the observations, where "faint" targets cannot be observed immediately after "bright" ones. All these aspects make the instrument ill-suited for the observations of faint targets.
\item \textbf{Thermal control.} The instrument does not include thermal control and the temperature
varies according to the external conditions and the variable power dissipated by the CCD.
The enclosure does not provide thermal insulation so the stability of the instrument is
low.

\item \textbf{Calibration}. There is no internal continuum lamp, flat fielding is obtained on the dome and it is a non-optimal and time-consuming process. A Thorium-Argon (ThAr) lamp is installed at the instrument front end for
non-simultaneous wavelength calibration.
\item \textbf{Acquisition.} The acquisition optics are made with a simple commercial achromatic doublet
so that the image quality is less than ideal, yet enough for centering the target and providing the basic
guiding.
\item \textbf{Instrument software.} There is no dedicated software; the science data acquisition is performed
with the commercial software Maxim DL.  
\item \textbf{Telescope.} PUCHEROS is installed at a $50$\,cm telescope and the small aperture of the telescope strongly
limits the brightness of the targets that can be observed. In addition, the site, only a few tens kilometers from downtown Santiago, is affected by strong light pollution and the seeing is rarely better than $2-3$\,arcsec. On the positive side however, the number of observable nights is above 60\% which is certainly above the average of a typical academic observatory at other sites in the world.
The scale of the telescope on its focal plane is $27.5$ \,arcsec\,mm$^{-1}$. The light of the stellar target is collected through a pinhole of $0.125$\,mm diameter
equivalent to an aperture of about $3.5$\,arcsec in the sky.
\item \textbf{Lack of remote observing.} PUCHEROS does not have a software system that allows remote observing, which limits the operation of the instrument to in person only. This is not an extreme difficulty for a telescope very near Santiago, where PUCHEROS is located, however, it becomes significantly more complex when the instrument is situated at a professional observatory, which is typically located far from urban areas. Therefore, remote observing is crucial for transforming PUCHEROS into a scientific competitive instrument.
\end{enumerate}

The performance achieved by PUCHEROS includes a limiting magnitude in V-band around $V=9$ and a precision on radial velocities of about 200\,m$\,$s$^{-1}$ \citep[measured with spectra from radial velocity (RV) standard stars LHS 249, HD 71479 and LTT 3283;][]{Vanzi2012}. This is good enough to follow up transient objects such as Novae or to study bright binary systems and perform academic exercises. In fact since its installation, PUCHEROS is routinely used as a teaching instrument in the courses of astronomy and it is also successfully used for research \citep[see i.e.][]{Chesenau2014,Izzo2015,Coronado2015,Bluhm2016,Arcos2018}.
 
\section{PUCHEROS+}\label{section: sec3 PUCHEROS+}

For all the aforementioned reasons, it seemed reasonable to upgrade the instrument, maintaining the original general idea but exploiting all its potential to reach more interesting scientific performances. This led to the development of the PUCHEROS+ spectrograph. 

This instrument is also a pathfinder for PLATOSpec, a high-resolution echelle spectrograph 
intended as a ground-based follow-up for planetary candidates from the TESS, PLATO, and ARIEL space missions \citep{Kabath2022PLATOSpec}. 
PUCHEROS+ allowed to test the telescope and all the associated systems to ensure stability and remote observing of the next instrument.

In this section, we present in detail how PUCHEROS+ overcomes all the limitations of the original instrument and supports the validation of the infrastructure necessary for PLATOSpec.\\

\subsection{The spectrograph}\label{subsection: sec3 spec}
The new instrument employs as the detector an Andor iKon-M 934 CCD, which is back-illuminated and operated at $-80^\circ$C with an A/D rate of $1$\,MHz. This detector provides excellent QE ($95$\% at $550$\,nm), low read noise (up to $2.6$\,e$^-$) and dark signal ($0.001$\,e$^-$\,pix$^{-1}$\,s$^{-1}$). The low temperature of operation removes completely any memory effect on the image. Heat extraction is performed by liquid closed cycle cooling which is highly efficient and enables precise temperature control throughout the entire instrument. The CCD fan only activates when there is an unexpected temperature increase due to possible problems in the liquid cooling. The shutter was removed from the detector to avoid the dissipation of a variable power; instead a smaller shutter is installed at the exit of the light injection system of the spectrograph. This shutter is bi-stable so that the amount of power dissipated is negligible. \\

The format of the CCD is $1K$ x $1K$ with pixels of $13.5\,\mu$m in size (in comparison with the $24\,\mu$m of the FLI camera of the original instrument). This makes the detector nearly half the size of the original one. To compensate for that, the entire original $f/3.5$ objective was replaced by a commercial photographic lens of $135$\,mm effective focal length and focal ratio $f/2$, which provides the image quality needed. 

\begin{figure}
    \centering
    \begin{subfigure}[b]{1\linewidth}
        \centering
    \includegraphics[width=1\linewidth]{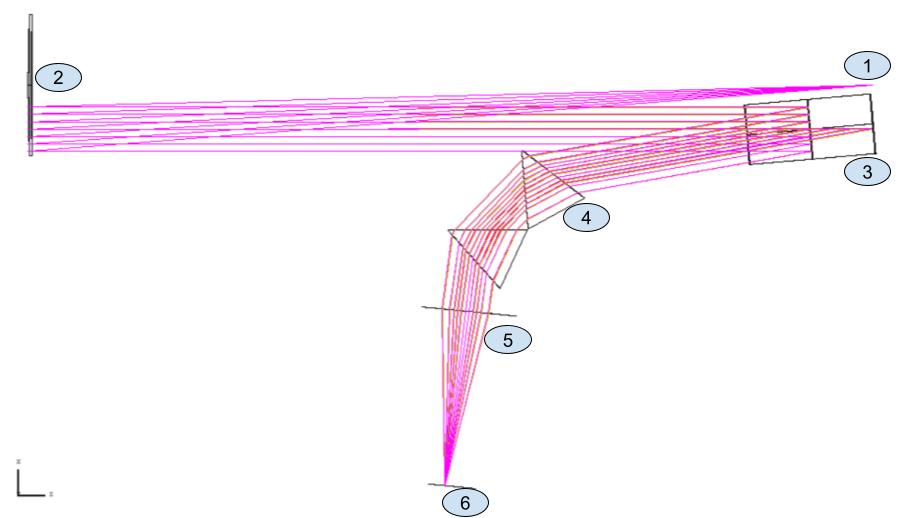}
    \caption{Optical design PUCHEROS+}
    \label{fig:pucheros_design}
    \end{subfigure}
    
    \begin{subfigure}[b]{1\linewidth}
        \centering
    \includegraphics[width=1\linewidth]{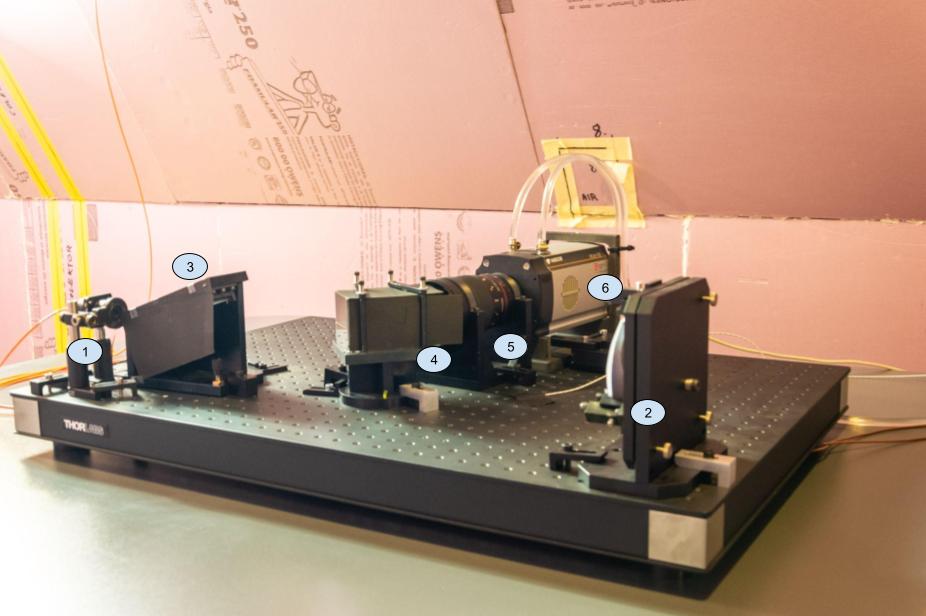}
    \caption{PUCHEROS+ installed at the ESO 1.52\,m telescope at La Silla observatory}
    \label{fig:pucheros+_installed}
    \end{subfigure}
    
    \caption{PUCHEROS+ optical design (a) and components (b): (1) Optical fiber input and shutter, (2) Collimator, (3) Echelle, (4) Prisms, (5) Objective (6) Detector}
    \label{fig:pucheros_design_installed}
\end{figure}

 The remaining opto-mechanical components of the spectrograph were kept unchanged. Identically to PUCHEROS, the $50\,\mu$m output of the fiber is magnified by a factor 4, which is done by a $12.5$\,mm focal length triplet lens, producing a $200\,\mu$m image. The resulting beam illuminates the spectrograph collimator with an f-number of approximately $f/18.5$. An image of the spectrograph optical design and assembly is shown in Figure \ref{fig:pucheros_design_installed}.\\

\subsection{Thermal control}\label{subsection: sec3 spec}

The spectrograph is hosted inside a thermally insulated enclosure and includes active thermal control. The enclosure consists of an aluminum structure of $30\times30$\,mm T-slot profiles, creating an inner insulated volume of $952\times$ $680\times430$\,mm. The sides are composed of $30$\,mm thickness, $30$\,kg\,m$^{-3}$ density polystyrene foam, externally covered by composite aluminum material (ACM) sheets of $3$\,mm thickness ($0.5$\,mm aluminum, $2$\,mm PE foam, $0.5$\,mm aluminum), and internally, covered with black suede fabric. The contact between the aluminum structure and the table is insulated by a $50\times 50$\,mm cross-sectional hard-foam. A temperature difference $\Delta T=1^\circ$C between inside and outside of the enclosure would produce a conduction exchange power of about $4$\,W.

To achieve the best possible stability inside the enclosure, we implemented three levels of control: instrument enclosure, instrument laboratory, and equipment room. The equipment room is the first stage; it is a space that contains the detector closed cycle chiller and the computers and electronic systems, the temperature of this space is maintained at $T=17 \pm{1}^\circ$C with an air conditioning (AC) equipment and a heater that are both actively controlled.

An air fan injects air from the equipment room into the instrument room and is activated when a temperature difference exceeding 1$^\circ$C is detected between the two spaces. To achieve this, two \textit{DHT11} sensors continuously monitor the temperature in each area.
The temperature in the enclosure is controlled with a BelektroniG Temperature Controller HAT that provides that provides a proportional–integral–derivative control (PID) over the voltage applied to an array of six $20$\,W heater resistors. These are connected in series on an aluminum plate below the instrument, the heat is transferred to the optical bench by air convection and radiative transfer. Temperature feedback is provided to the controller by one RTD \textit{Pt1000} sensor with $0.01$K resolution. The PID sampling rate is of $10$\,Hz, and the values for the PID constants were tuned with empirical tuning methods.  This setup provides a thermal stability of $\Delta T=0.2$\,K on the optical bench. \\

\subsection{Telescope front-end}\label{subsection: sec3 tfe}
\begin{figure}
    \centering
    \includegraphics[width=1\linewidth]{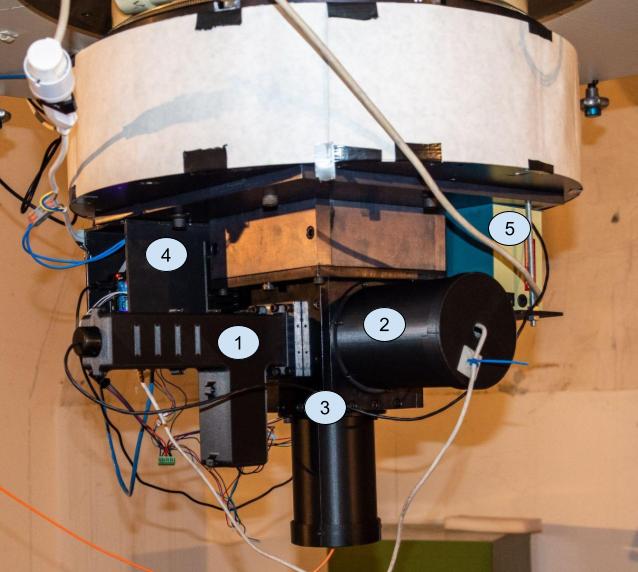}
    \caption{Front-end of PUCHEROS+ installed, with its components: (1) Calibration system, (2): Acquisition and guiding system, (3) Fiber Injection system, (4): Electronic Box, (5) Thorium-Argon lamp power source.}
    \label{fig:Interface}
\end{figure}

PUCHEROS+ was first tested at the ESO 1.52\,m in La Silla observatory. The telescope is a $\sim f/15$ Cassegrain with a scale at the focal plane of $9.1$ arcsec\,mm$^{-1}$. A pinhole of $0.175$\,mm gives an aperture on the sky of $1.6$\,arcsec. The telescope front-end connects and provides light injection into the fiber, acquisition of the target, guiding, and calibration of the spectrograph. An image of the entire front-end is shown in Figure \ref{fig:Interface}.

\begin{figure}
    \centering
    \begin{subfigure}[b]{1\linewidth}
        \centering
    \includegraphics[width=1\linewidth]{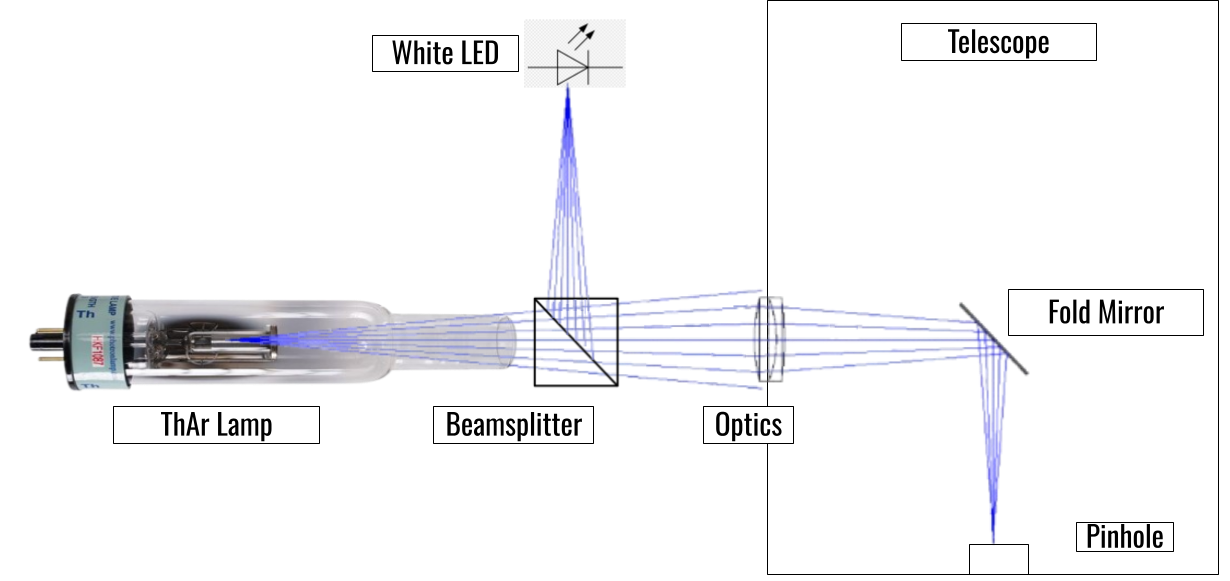}
    \caption{Conceptual design: Thorium-Argon lamp for spectral calibration and white LED for flat-field, passing through a $90/10$ beamsplitter and focusing on the pinhole, through a fold mirror.}
    \label{fig:CalibrationSystemDesign}
    \end{subfigure}
    
    \begin{subfigure}[b]{1\linewidth}
        \centering
    \includegraphics[width=1\linewidth]{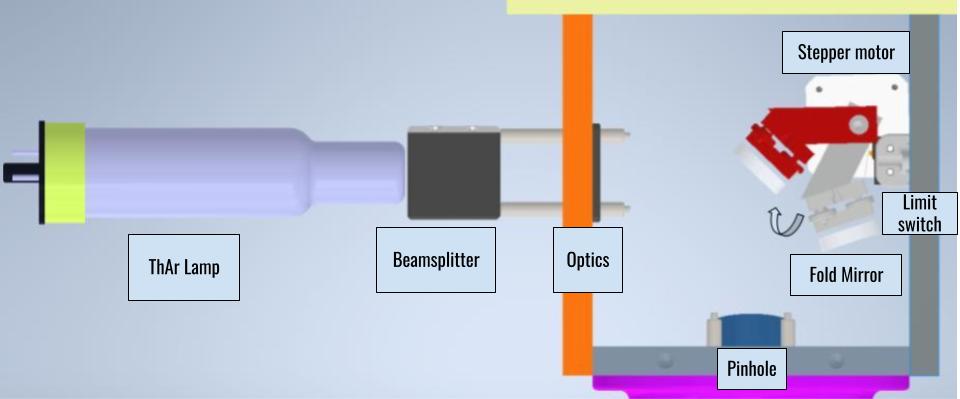}
    \caption{Mechanical design: Thorium-Argon lamp illuminating the pinhole over a moving fold mirror, controlled by a stepper motor and a limit switch.}
    \label{fig:CalibrationMechanicalDesign}
    \end{subfigure}
    
    \begin{subfigure}[b]{1\linewidth}
        \centering
    \includegraphics[width=1\linewidth]{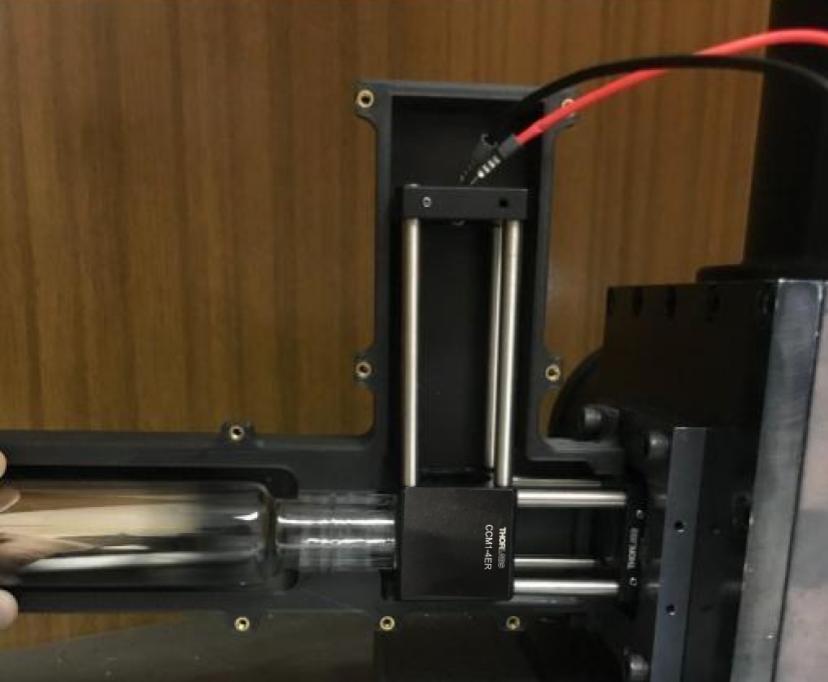}
    \caption{Components: Thorium-Argon lamp on the left, white LED at the top, beamsplitter at the bottom, and optics on the right.}
    \label{fig:CalibrationComponents}
    \end{subfigure}
    
    \caption{Calibration System}
    \label{fig:CalibrationSystem}
\end{figure}

The calibration system is attached to the front-end cube, so it moves with the telescope. This system consists of two calibration sources: a Thorium-Argon lamp for spectral calibration and an LED for flat-fielding and order identification; these calibration sources illuminate a $90/10$ beamsplitter, respectively, to further go through a $75$\,mm lens, to form an image on the $175\,\mu$m pinhole on the telescope focal plane. The calibration optics were chosen to get the whole system fixed to a $130 \times 130 \times 130$\,mm aluminum cube. In order not to interfere with the path of the light of the stars when observing, the light from calibration sources is injected perpendicularly to the axis of the telescope, so a moving fold mirror is necessary to redirect this light to the pinhole. The calibration system is shown in Figure \ref{fig:CalibrationSystem}, illustrating both its design and components. The calibration sources, the beamsplitter, and the optics are located outside the cube. Inside the cube, there are the fold mirror, a stepper motor that controls its movement, a limit switch defining the \textit{REST} position of the mirror (observing position), the acquisition mirror, the acquisition objective, and the pinhole.

\begin{figure}
    \centering
    \includegraphics[width=1\linewidth]{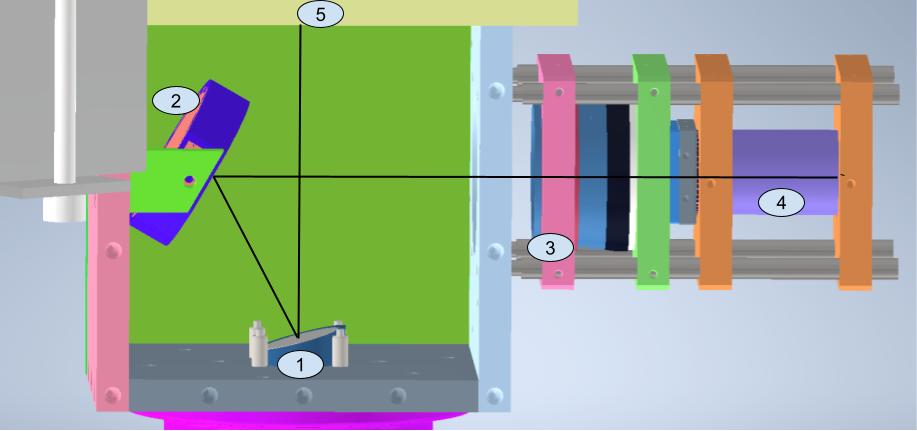}
    \caption{Acquisition and guiding system optomechanical design (optical axis shown): Pinhole mirror (1), fold mirror (2), objective (3), detector (4) and telescope (5). Note that this figure shows the same cube than Figure \ref{fig:CalibrationComponents} but in perpendicular direction.}
    \label{fig:AcqGuidAxis}
\end{figure}

The acquisition and guiding system consists on a photographic objective of $35$\,mm focal length and $f/1.2$; a $50.8$\,mm diameter fold mirror and an Imaging Source DMK 27AUR0135 camera, with $1280 \times 960$ pixels of $3.75\,\mu$m diameter. The acquisition system images a $25.4$\,mm reflective surface around the pinhole (referred to as the pinhole mirror) which, given the telescope plate scale, gives a field of view of $3.85$\,arcmin and a scale in the acquisition camera of $0.14$\,arcsec\,pix$^{-1}$. 
The optomechanical design of the acquisition and guiding system can be seen in Figure \ref{fig:AcqGuidAxis}, where we can see the optical axis of the light beam from the telescope that reflects off the pinhole mirror, then gets redirected by the fold mirror, and it is finally focused by the objective onto the acquisition detector.
With a maximum frame rate of $60$ FPS, it is smooth enough to provide good guiding.

The injection-to-the-fiber system was made with a commercial triplet with 12.5\,mm of focal length, that takes the light from the telescope at $\sim f/15$ and injects it into a $50\,\mu$m fiber at $f/4.3$.

\subsection{Software systems and data processing}

\subsubsection{Software design and implementation}
We developed a dedicated instrument Software for PUCHEROS+ which provides control of the scientific CCD, acquisition camera and front-end. 
The design and implementation of this software presented some interesting challenges, such as the need to communicate and coordinate tasks between the front-end and the spectrograph. This requires the software to be cross-platform and support multiple programming languages. To meet these requirements, we decided to use the Internet Communications Engine (ICE) \citep{henning2003distributed}, an open-source RPC framework that provides the necessary tools.

To control the devices described in section \ref{subsection: sec3 tfe}, a software subsystem called Instrument Control Software (ICS) was developed. The ICS subsystem is composed of a module called Lamp, responsible for turning the white LED on and off; a module called Mirror, responsible for inserting or removing the calibration mirror; and a module called Relay, responsible for turning the Thorium-Argon lamp on and off.

The subsystem Detector Control System (DCS) was developed to manage both the acquisition detector and the scientific detector. It includes two modules: TDCS, which controls the Imaging Source DMK 27AUR0135 camera, and SDCS, which manages the Andor iKon-M 934.

The OS module manages the calibration procedures of the spectrograph and handles the acquisition and storage of scientific data. The Graphical User Interface (GUI) module provides the interface through which the observer monitors the spectrograph status and issues operational commands. The GUI module interfaces with the OS module to perform calibration, data acquisition, and status monitoring tasks.

\begin{figure}
    \centering
    \includegraphics[width=1\linewidth]{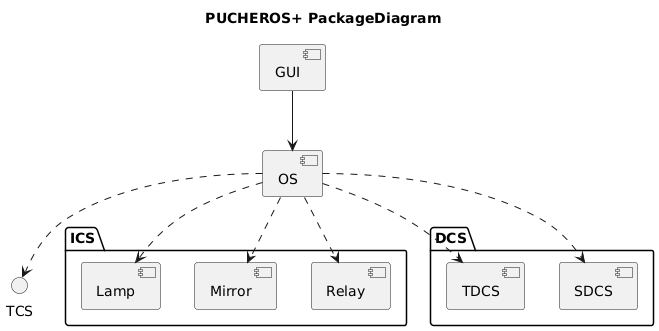}
    \caption{Pucheros+ Package Diagram.}
    \label{fig:packdia}
\end{figure}

Finally, the Telescope Control System (TCS) acts as a client that retrieves key telescope parameters, including status, coordinates, velocities, and dome positions. This information is used to populate the corresponding metadata fields in the scientific FITS files. A diagram of the software system can be seen in Figure \ref{fig:packdia}.

The Andor detector is connected by an USB3 cable to a HPE Proliant MicroServer Gen10 Plus (16GB of RAM). The control of the acquisition camera and front-end is done through a Raspberry Pi 4 (named \textit{rpi1}), which also controls the relays for the calibration lamps supplies and a NEMA17 stepper motor of the calibration mirror. The software for the acquisition camera is based on the \textit{tiscamera} SDK provided by Imaging Source. The air fan that injects air to the spectrograph's room is controlled by another Raspberry Pi 4 (named \textit{rpi2}) for reading the \textit{DHT11} sensors and an Arduino UNO to control a relay to give power to the fan.

\subsubsection{Remote observing}
As seen in section \ref{section: sec2 PUCHEROS}, one of the main requirements for PUCHEROS+ was the possibility of remote observing from Santiago de Chile and from various European locations. Therefore, we developed in-house control software similar to the one used for OES spectrograph at Perek telescope \citep{Kabath2020}. The routines controlling the pointing and guiding of the telescope are written in \texttt{python} and based on the Astropy library \citep{greenfield2013astropy}. 

The observing process requires the observer to first acquire calibration frames through the CCD GUI. Then, to begin stellar observations, the observer points the telescope using the telescope GUI, either by entering the star ID (recognized by the SIMBAD Astronomical Database) or by manually inputting the right ascension (RA) and declination (DEC) coordinates. The acquisition and guiding are performed in GUIs by the observer who can center the star on the fiber, and then an automatic routine performs the guiding. Finally, the observer acquires the science frames using the CCD GUI. Due to the non-simultaneous calibration of the spectrograph, a spectral lamp calibration frame must also be taken after every science frame.

The software runs independently in Chile. In case of connection failure to the remote location, the guiding is not influenced. Furthermore, automatic scripting can provide long time series and thus connection problems are not interfering with operations until human interaction such as during the acquisition is required.

The telescope network provides remote access via VPN to the instrument server, \textit{rpi1} and \textit{rpi2}. Uninterruptible power supply is guaranteed by the UPS Observatory power network.

\subsubsection{Data processing}

The spectra are processed with a dedicated automated pipeline based on \texttt{CERES} \citep{brahm2017ceres} package. The new \texttt{CERES} version (called \texttt{ceres+}) is an object oriented \texttt{python} package that contains a collection of routines that allows one to generate automated data processing pipelines for different types of echelle spectrographs. In the case for PUCHEROS+, the dedicated \texttt{ceres+} pipeline performs all processing steps required to generate wavelength calibrated output spectra and radial velocity measurements.
The pipeline classifies all (calibration and scientific) raw images based on the information available in the headers. After this classification, the pipeline generates master calibration frames (\textit{masterbias}, and \textit{masterflat}).
All echelle orders are identified and traced by the pipeline from the \textit{masterflat}.
After this, the scientific images are corrected by scattered light contamination. The spectra associated to the scientific images are generated using an optimal extraction algorithm presented in \citet{marsh}. 

For the wavelength calibration (ThAr) spectra, the pipeline uses a simple extraction algorithm.
The pipeline computes the wavelength solution for all extracted ThAr spectra by measuring the exact pixel position (at sub-pixel level) of a predefined set of prominent Th emission lines whose wavelengths are known from \citet{lovis}. The pipeline fits a two-dimensional Chebyshev polynomial of the same form as presented in \citet{brahm2017ceres}. The degree of the polynomial is of $3$ in pixel position and $5$ in echelle order number. The pipeline takes the first ThAr spectrum of a given night as the zero point, and then the pipeline computes the velocity drifts between all the other ThAr spectra and the first one. The typical achievable radial velocity precision for an RV drift calculation is $\sim 8\,$m\,s$^{-1}$. Then, with this information in hand, the pipeline is able to estimate the instrumental drift at any time in the night using a time linear interpolation of the drifts computed with the ThAr spectra.

The \textit{masterflat} spectrum associated to a given night is also extracted and used to partially correct the blaze function of the scientific spectra using a simple division. The scientific spectra are further continuum normalized using an iterative polynomial fit.

The barycentric velocity correction is computed using the algorithm incorporated into the \texttt{astropy} package \citep{greenfield2013astropy}.

The pipeline also incorporates functions to compute the cross-correlation function of each continuum normalized spectrum. The templates are the same binary masks used by the pipeline of the HARPS spectrograph. The cross-correlation peak is modeled with a simple Gaussian function and the mean of the Gaussian corresponds to the reported radial velocity of the star. The pipeline also computes the bisector span and the full width at half maximum of the cross-correlation peak.

\section{Results}\label{section: sec4 Results}
PUCHEROS+ was installed at the ESO 1.52\,m telescope in La Silla in September 2022. In this section, we present the commissioning results, including performance of the detector, efficiency and signal-to-noise ratio obtained from the observations, stability of the temperature and radial velocity precision. 
From the reduced spectra, generated by the pipeline, we can obtain the flux in Analog-to-Digital Units (ADUs) for each order, the radial velocity of each observed star and the instrumental drift measured through the night in the calibration spectra.

\subsection{Image quality}
The size of the image of the fiber on the detector is determined by the demagnification applied to the $200\,\mu$m input fiber image by the spectrograph optics. This factor is defined by the focal lengths of the collimator and the objective ($M = 1/4.8$). Considering the detector pixel size, the resulting image should be of $3.1$ pixels in diameter. 

\begin{figure}
    \centering
    \begin{subfigure}[b]{1\linewidth}
        \centering
    \includegraphics[width=9cm]{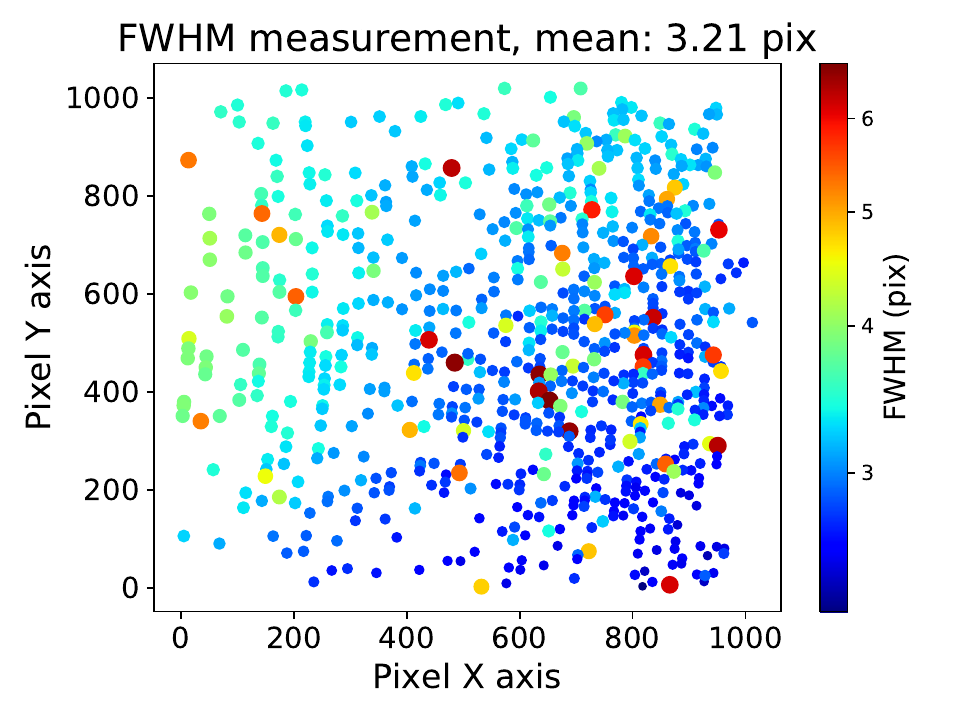}
    \caption{2D FWHM of fiber images obtained illuminating the spectrograph with ThAr lamp. The FWHM is shown on logarithmic scale in both the size and color of each fiber image.
    The remaining fiber images with large FWHM values are mostly due to errors in the source extraction analysis, specifically, cases where two or more emission lines are incorrectly detected as a single source.}
    \label{fig:2D FWHM}
    \end{subfigure}
    
    \begin{subfigure}[b]{1\linewidth}
        \centering
    \includegraphics[width=9cm]{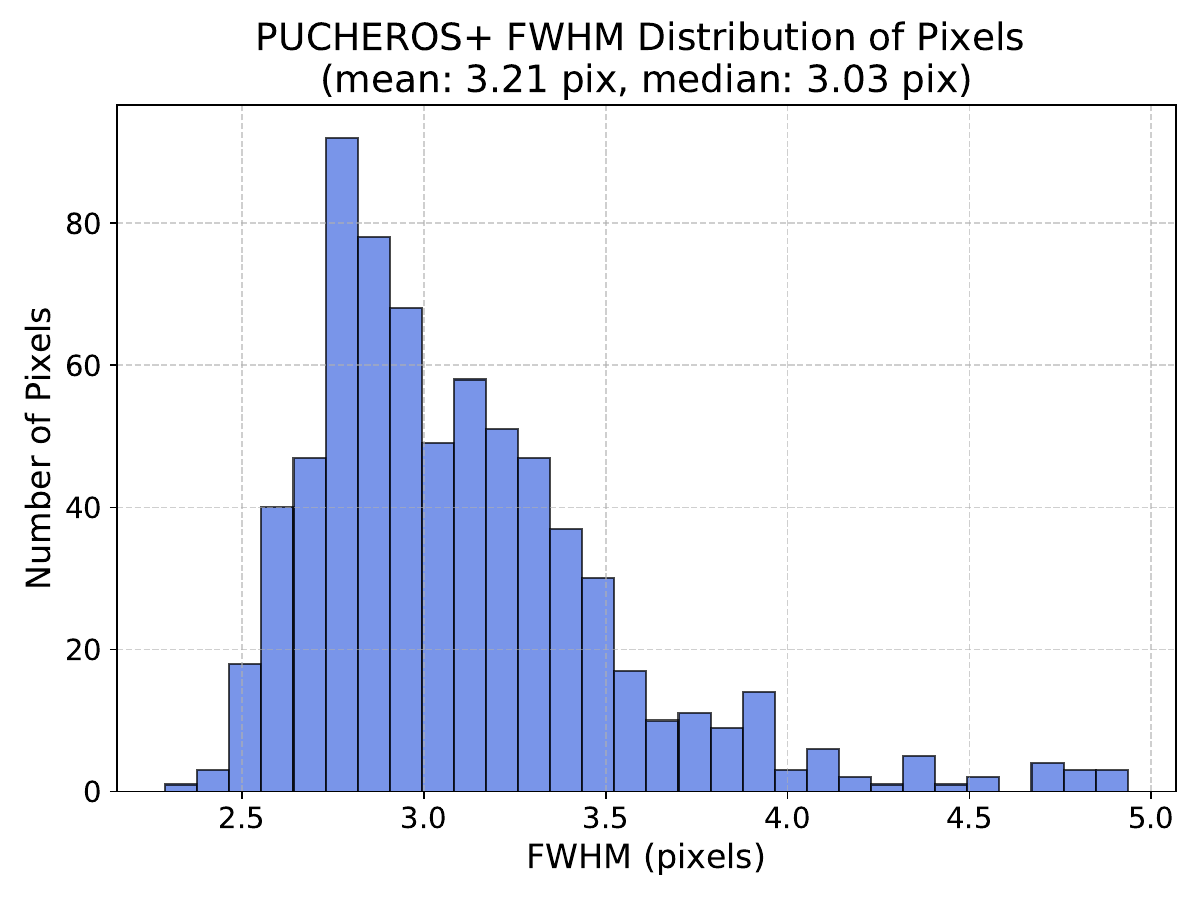}
    \caption{Histogram of 2D FWHM of illuminated fiber image.}
    \label{fig:Histogram_FWHM}
    \end{subfigure}
    
    \caption{Measurement of the FWHM in a ThAr spectrum of PUCHEROS+.}
    \label{fig:ImgQuality}
\end{figure}

Thus, we can measure the image quality achieved by the spectrograph by measuring the full-width-half-maximum (FWHM) of a Gaussian fit of the fiber image in the detector. We can perform this measurement by analyzing the images of the emission lines of a ThAr spectrum using the source detection algorithm implemented in SExtractor (v2.12.4) \citep{bertin1996sextractor}. Detection and analysis thresholds were set to ensure a minimum pixel signal-to-noise ratio (SNR) of 50. Using this method, more than $770$ objects were extracted and analyzed. 
The results, after removing the top $5$\% of the data points (outliers caused by saturated lines), are shown in Figure \ref{fig:2D FWHM}, where the mean FWHM is $3.2$ pixels. Figure \ref{fig:Histogram_FWHM} shows a histogram of the 2D FWHM values, indicating that a $95$\% of the fiber images detected in the spectrum have a FWHM between $2.5$ and $4$ pixels.

\begin{figure}
    \centering
    \includegraphics[width=1\linewidth]{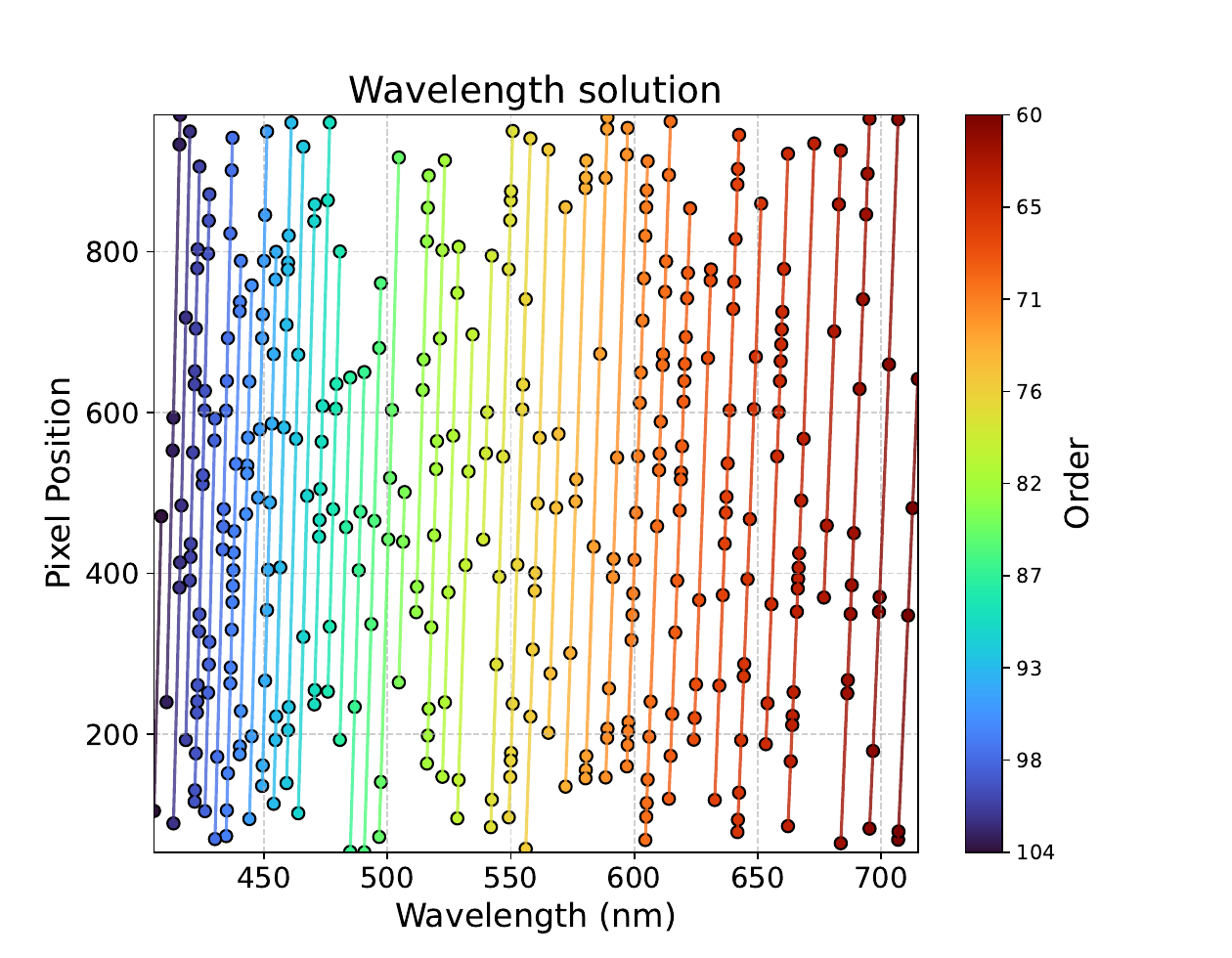}
    \caption{Representation of the wavelength solution of a ThAr spectrum of PUCHEROS+.}
    \label{fig:WavSol}
\end{figure}

\begin{figure}
    \centering
    \includegraphics[width=1\linewidth]{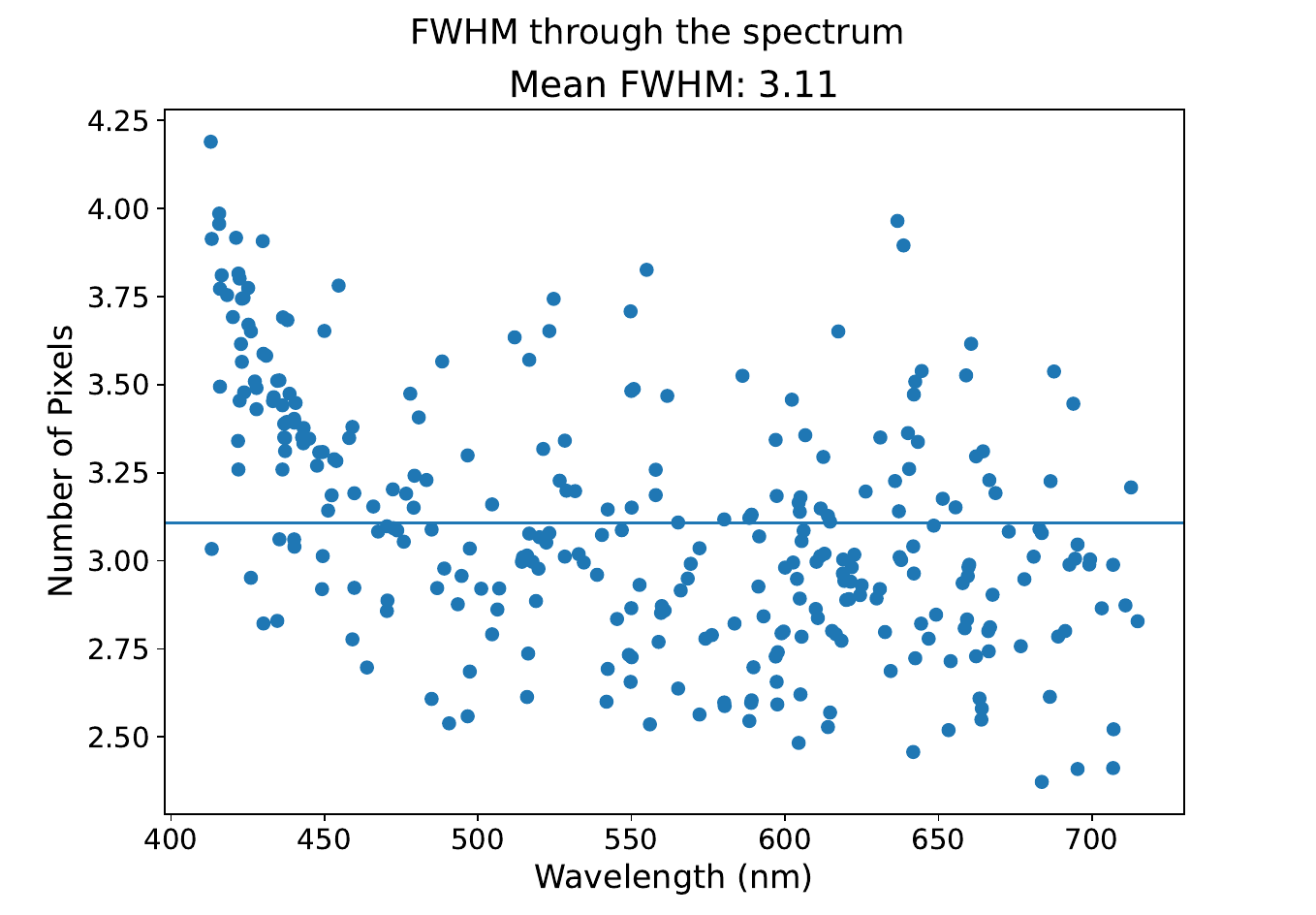}
    \caption{Measured spectral sampling of a ThAr spectrum of PUCHEROS+.}
    \label{fig:Sampling}
\end{figure}

\begin{figure}
    \centering
    \includegraphics[width=1\linewidth]{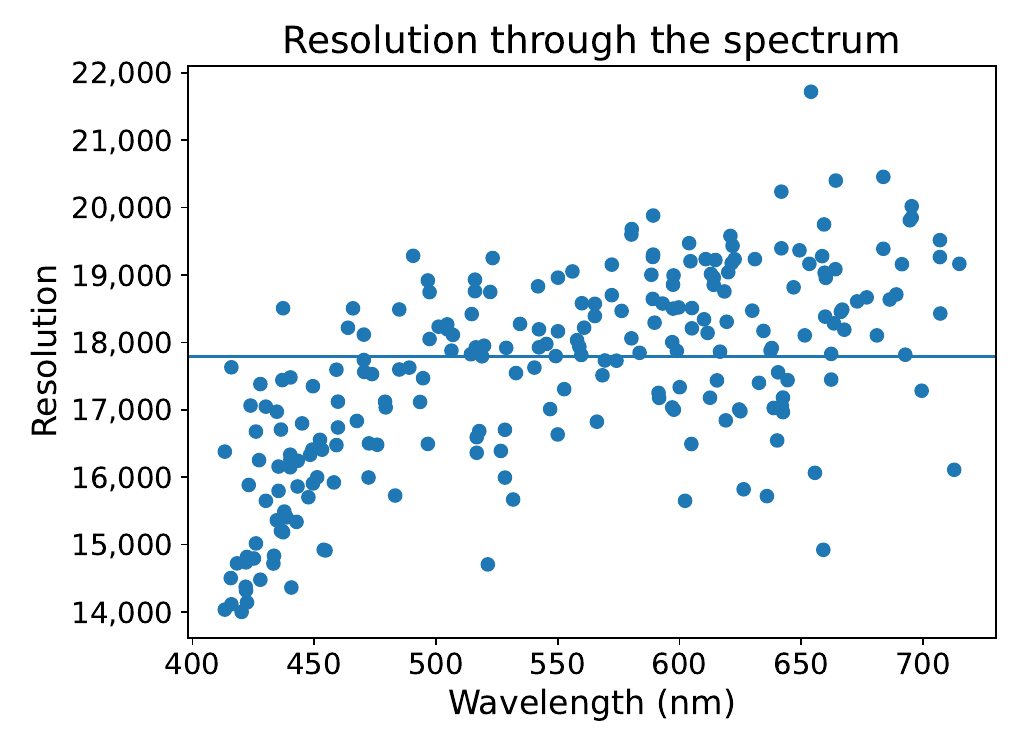}
    \caption{Measured resolution of a ThAr spectrum from PUCHEROS+. The horizontal line represents the nominal value of the resolution.}
    \label{fig:Resolution}
\end{figure}

In addition to the above, for each ThAr calibration the pipeline provides 1D extracted spectra and the corresponding wavelength solution. The typical wavelength solution is shown in Figure \ref{fig:WavSol}. From the extracted spectra it is possible to calculate the 1D-FWHM (measured in pixels) of the detected emission lines and obtain the spectral sampling through the spectrum, as can be seen in Figure \ref{fig:Sampling}. The average spectral sampling is $3.1$ pixels, consistently with the expectation.
Thus, we can calculate the resolution of the spectrograph.
The theoretical spectral resolution of the spectrograph, given by the focal length of the objective, the size of the image of the fiber and the blaze angle of the echelle, is $R = 17,793$. Figure \ref{fig:Resolution} shows the spectral resolution across the detector, which is fully in agreement with the value indicated above. 

\subsection{Efficiency, SNR, Exposure time calculator}

To estimate the throughput of the system, first we measured the efficiency of each single optical component in the lab {using a HeNe laser ($633$\,nm) and a photometer. Table \ref{tab:efficiency} shows the values obtained, which provide a theoretical value of about $33$\% end-to-end spectrograph efficiency. The average efficiency of the fiber link can be estimated at about $60$\% including fiber attenuation and focal ratio degradation as estimated by laboratory measurements. For the telescope optics we consider an overall efficiency of about $70$\%. Based on this we can expect a maximum total on-sky efficiency of no more than $13-14$\%. 

Observations of standard stars were performed during approximately four months, reaching $597$ observations, obtaining an average efficiency of approximately $5$\%. The results, that are filtered by airmass $<1.5$, are presented in Figure \ref{fig:efficiency_frontend_old}. Some observations indeed reach the theoretical efficiency; however, there is also a large dispersion which can be attributed to pinhole losses, changing air-mass and atmospheric transparency. In particular, the aperture of $1.6$\,arcsec in the sky produces pinhole loss when seeing is less than optimal. For that reason, observations with efficiencies below $3$\% were excluded from the analysis, assuming that they were affected by bad seeing conditions.
\begin{table}
\caption{Theoretical efficiency of optical components}
\label{tab:efficiency}
\begin{tabular}{ll}
\textbf{Component}                            & \textbf{Efficiency}                           \\
\hline                                      
\textit{Collimator}                           & 85\%                                          \\
\textit{Echelle}                              & 50\%                                          \\
\textit{Prisms}                               & 93\%                                          \\
\textit{Objective}                            & 93\%                                          \\
\textit{CCD}                                  & 90\%                                          \\ \hline
\multicolumn{1}{l}{\textit{\textbf{Total}}} & \multicolumn{1}{l}{\textbf{33\%}} \\ \hline
\end{tabular}
\end{table}

\begin{figure}
    \centering
    \includegraphics[width=1\linewidth]{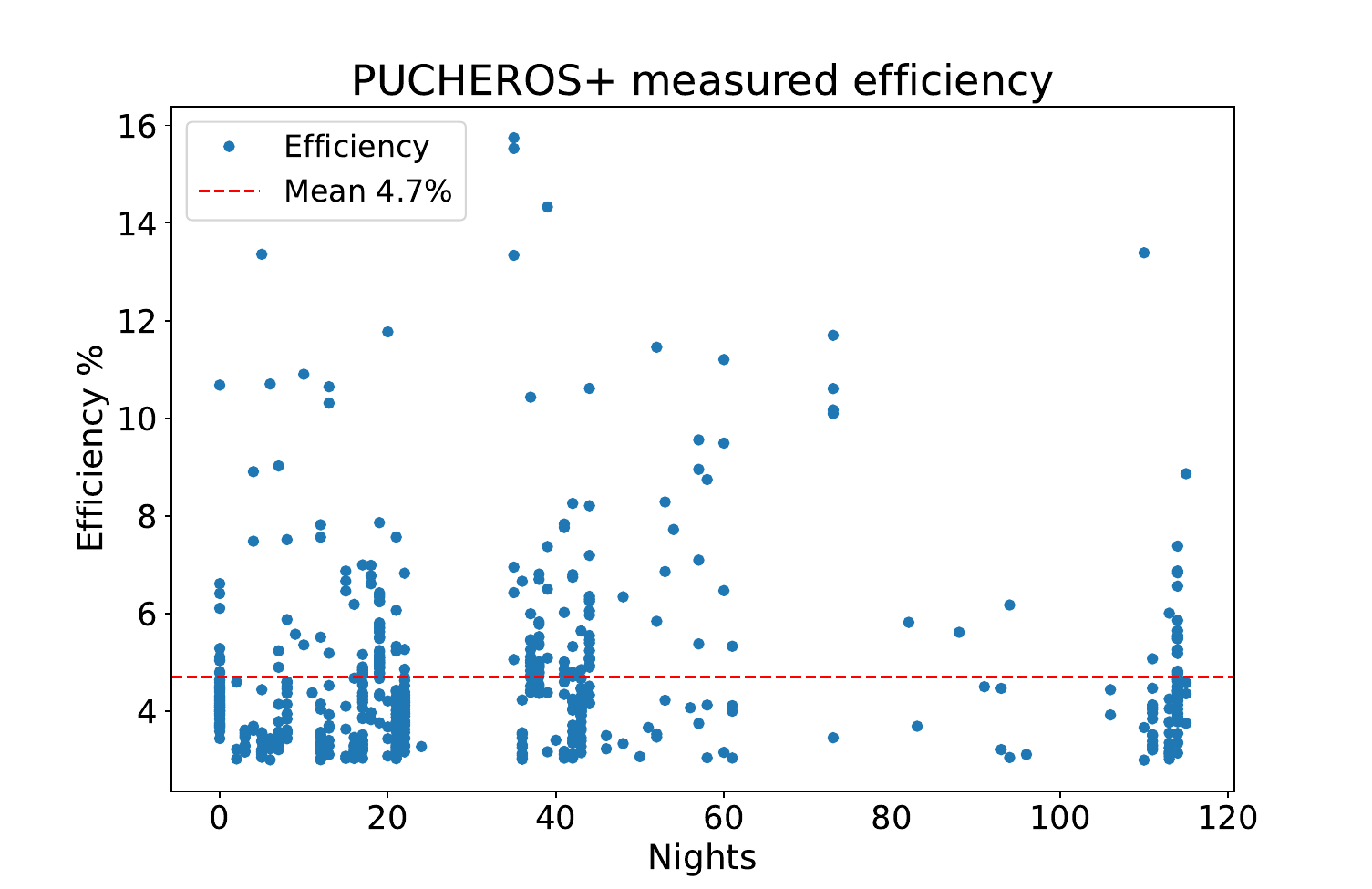}
    \caption{Efficiency of PUCHEROS+ over four months of observations, after filtering out observations with airmass $> 1.5$ and efficiency $< 3$\%.}
    \label{fig:efficiency_frontend_old}
\end{figure}

After setting the value of the efficiency of the system we can develop an exposure time calculator (ETC), estimating the signal-to-noise ratio (SNR) versus exposure time and \textit{V}-magnitude. Figure \ref{fig:ETC_frontend_old} shows the calculated SNR curves from \textit{V}-magnitude $6$ to $12$ and exposure time between $120$ and $1{,}800$ seconds along with a set of measured data points. The SNR of the observations was estimated by the \texttt{ceres+} pipeline in the region of the magnesium triplet ($\sim$517\,nm). We can notice that the system formed by the telescope, front-end and instrument, can achieve SNR of approximately $16$ for a $12$ \textit{V}-magnitude star in $1{,}800$ seconds exposure time. The differences between the calculated SNR curves and the pipeline estimated SNR (dots in Figure \ref{fig:ETC_frontend_old}) are due mainly to observational innacuracies.The curves are based on an average system efficiency of approximately $5$\%, which was adopted as the nominal on-sky efficiency. However, individual observations were subject to varying conditions, particularly changes in seeing, which have a more pronounced impact on fainter targets.
\begin{figure}
    \centering
    \includegraphics[width=1\linewidth]{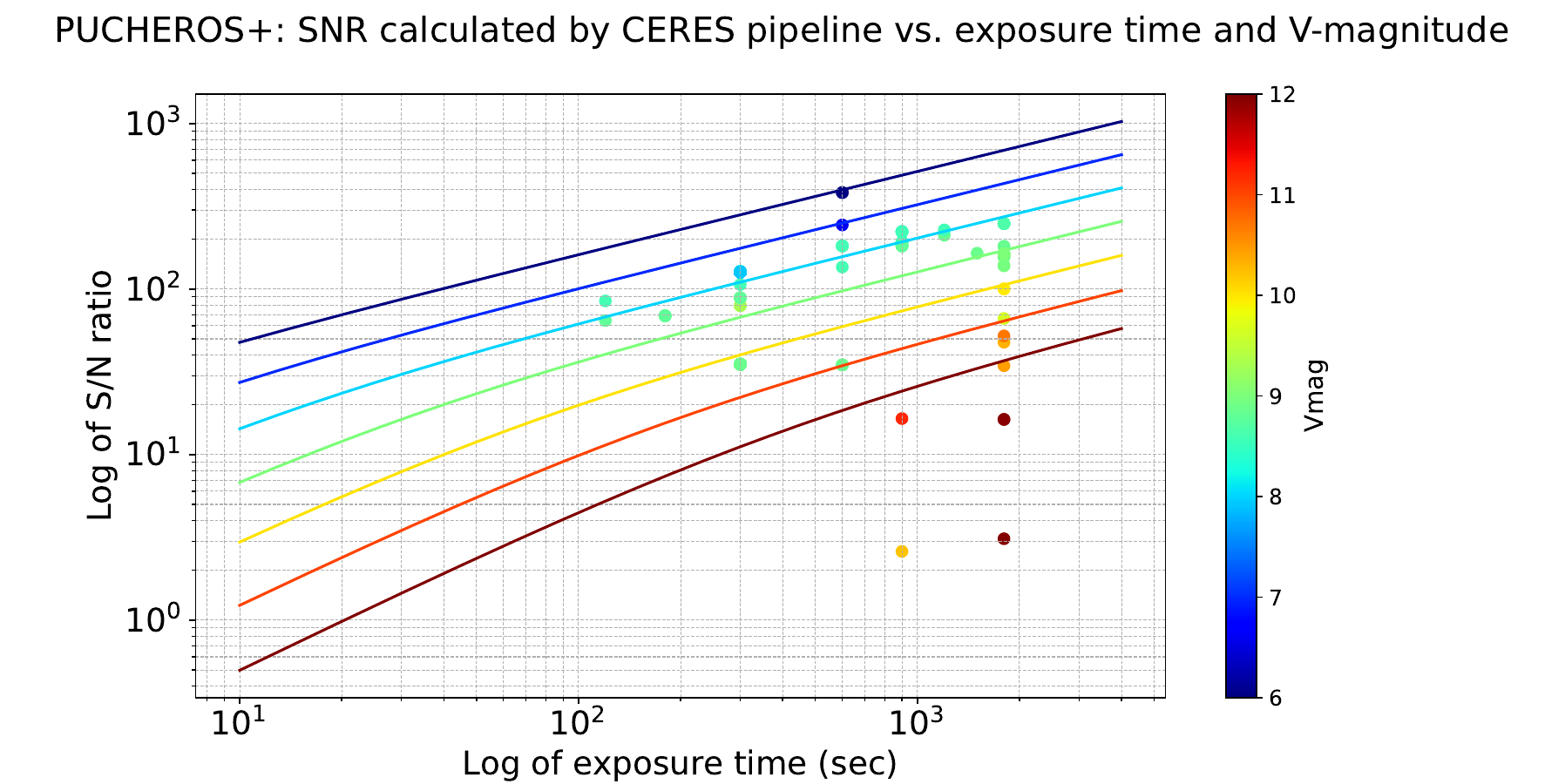}
    \caption{Exposure Time Calculator (ETC) results for PUCHEROS+ over four months of observations, after filtering out observations with airmass $> 1.5$ and efficiency $< 3$\%. Dots represent the pipeline-calculated SNR for each observation, while lines indicate the expected SNR as a function of exposure time for different V magnitudes.}
    \label{fig:ETC_frontend_old}
\end{figure}

\subsection{Temperature Stability}
As described in section \ref{section: sec3 PUCHEROS+}, {the thermal control provides an overall temperature stability of $\Delta=0.2$\,K or less. In Figure \ref{fig:temp-stability_1} we present the temperature evolution over one entire month, the temperature standard deviation is $\sigma = 60$\,mK in the optical bench, six times less than the variation of the instrument room. If we look closely at one-week data, where the room temperature was more stable, we can see that the thermal control system reaches variations of $\sigma = 40$\,mK, while the room varies three times more (Figure \ref{fig:temp-stability_2}). Temperature data was taken using high-accuracy RTD sensors \textit{Pt100} through an OMEGA OM-CP-QUARTD datalogger, with 10\,mK resolution. These monitoring sensors were independent from \textit{Pt1000} control sensor on the HAT controller, mentioned in section \ref{section: sec3 PUCHEROS+}.\\

\begin{figure}
    \centering
    
    \begin{subfigure}[b]{1\linewidth}
        \centering
    \includegraphics[width=1\linewidth,height=6cm]{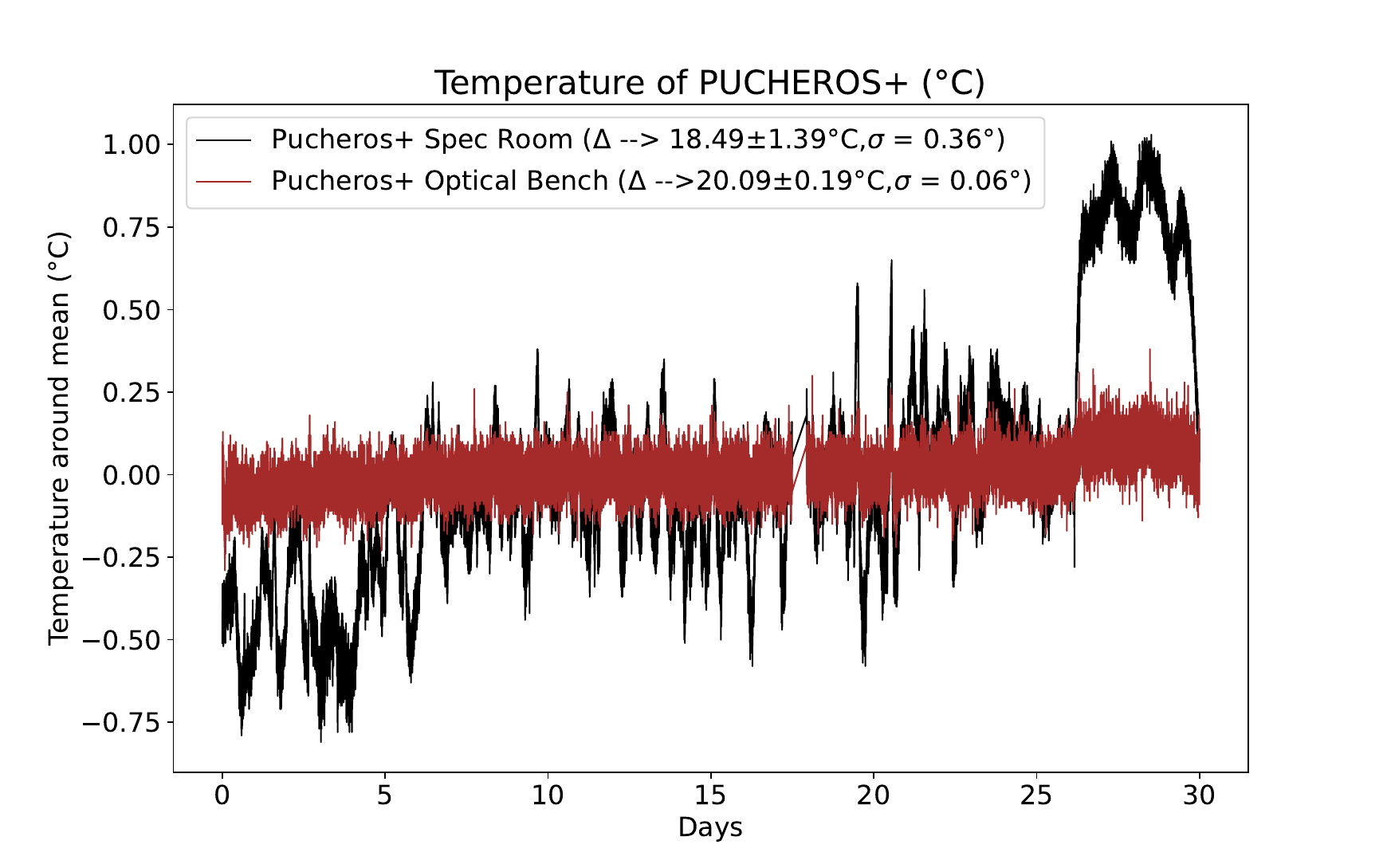}
    \caption{Over 1 month.}
    \label{fig:temp-stability_1}
    \end{subfigure}
    
    \begin{subfigure}[b]{1\linewidth}
        \centering
    \includegraphics[width=1\linewidth,height=6cm]{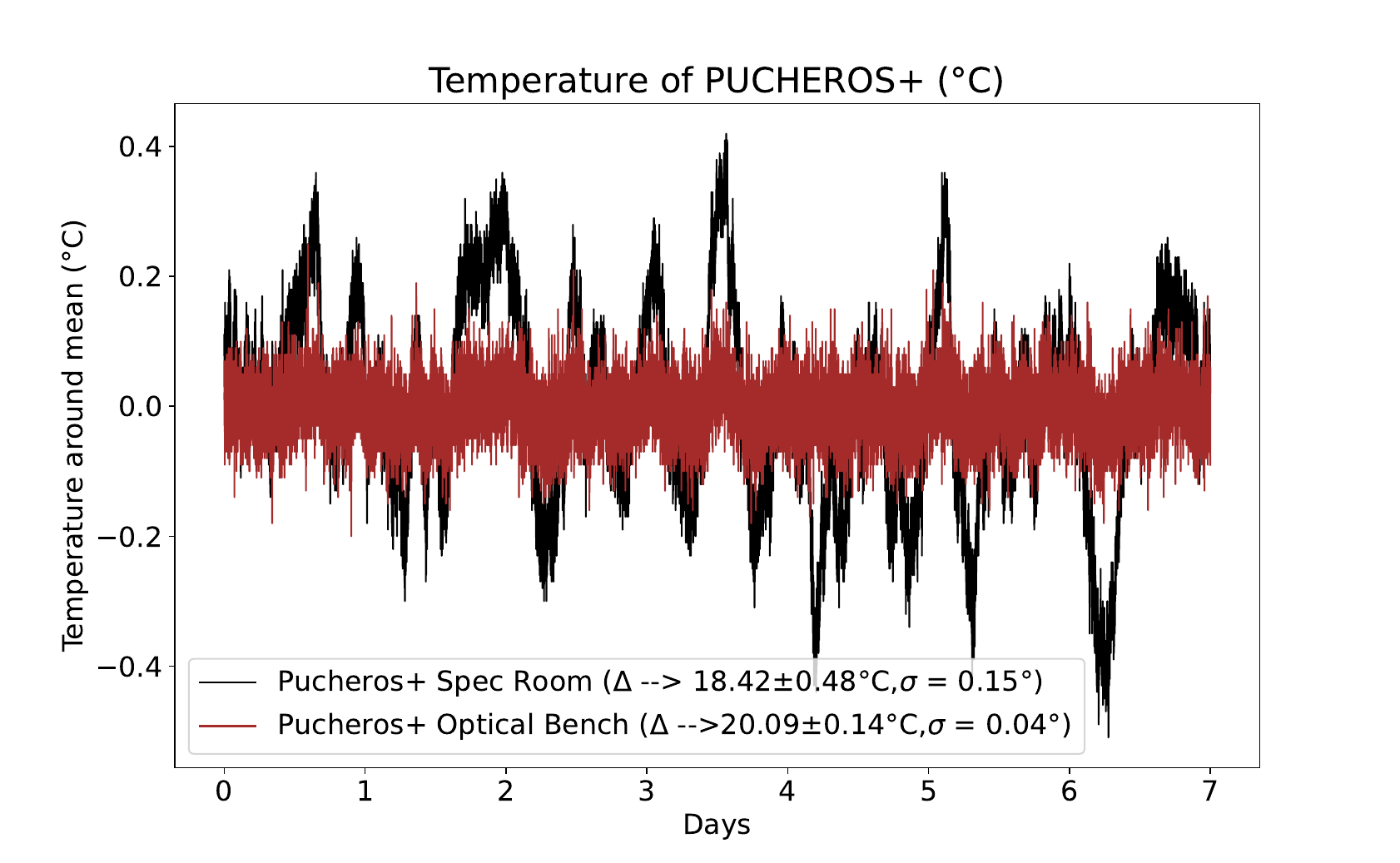}
    \caption{Over a week.}
    \label{fig:temp-stability_2}
    \end{subfigure}
    
    \caption{Temperature deviation curves for spectrograph Room (black curve) and optical bench temperature (brown curve), which is the actively temperature controlled surface, where a stability (range) of less than 0.2K is achieved for 30 days and less than 0.15K for 7 days. These statistics are after cleaning $10$\% outliers.}
    \label{fig:TempStability}
\end{figure}

\subsection{Radial velocity (RV) Stability}
The absence of a simultaneous wavelength calibration spectrum is a limitation to reach precision RV measurements. Calibration spectra must be acquired before and after each observation to ensure the best precision. To assess the RV precision achievable with PUCHEROS+, we analyze the behavior of the instrumental drift measured with the reference spectra of the ThAr lamp. For this purpose, we ran a long ThAr calibration sequence, taking a spectrum every $10$ minutes, for 10 hours approx. We divide these spectra into two groups evenly distributed in time and call them "calibration" and "science" (with the 'science' group simulating actual scientific observations). We then interpolate the "calibration" results and compare them with the values obtained from the "science" spectra to predict a precision limit for this type of observation, as the difference between the two. By this method, we can simulate the expected precision on exposures of 10, 20 or 30 min (typical scientific exposure times). The results are presented in Figure \ref{fig:MaximumPrecision}, where the maximum achievable precision  for this method (root-mean-square (RMS) of the residuals) is around $23.3$\,m\,s$^{-1}$, when simulating 10 minutes observations, and $34.6$\,m\,s$^{-1}$ for $30$ minutes observations, both in a $10$ hour ThAr sequence. 

\begin{figure}
    \centering
    \begin{subfigure}[b]{1\linewidth}
        \centering
    \includegraphics[width=1\linewidth]{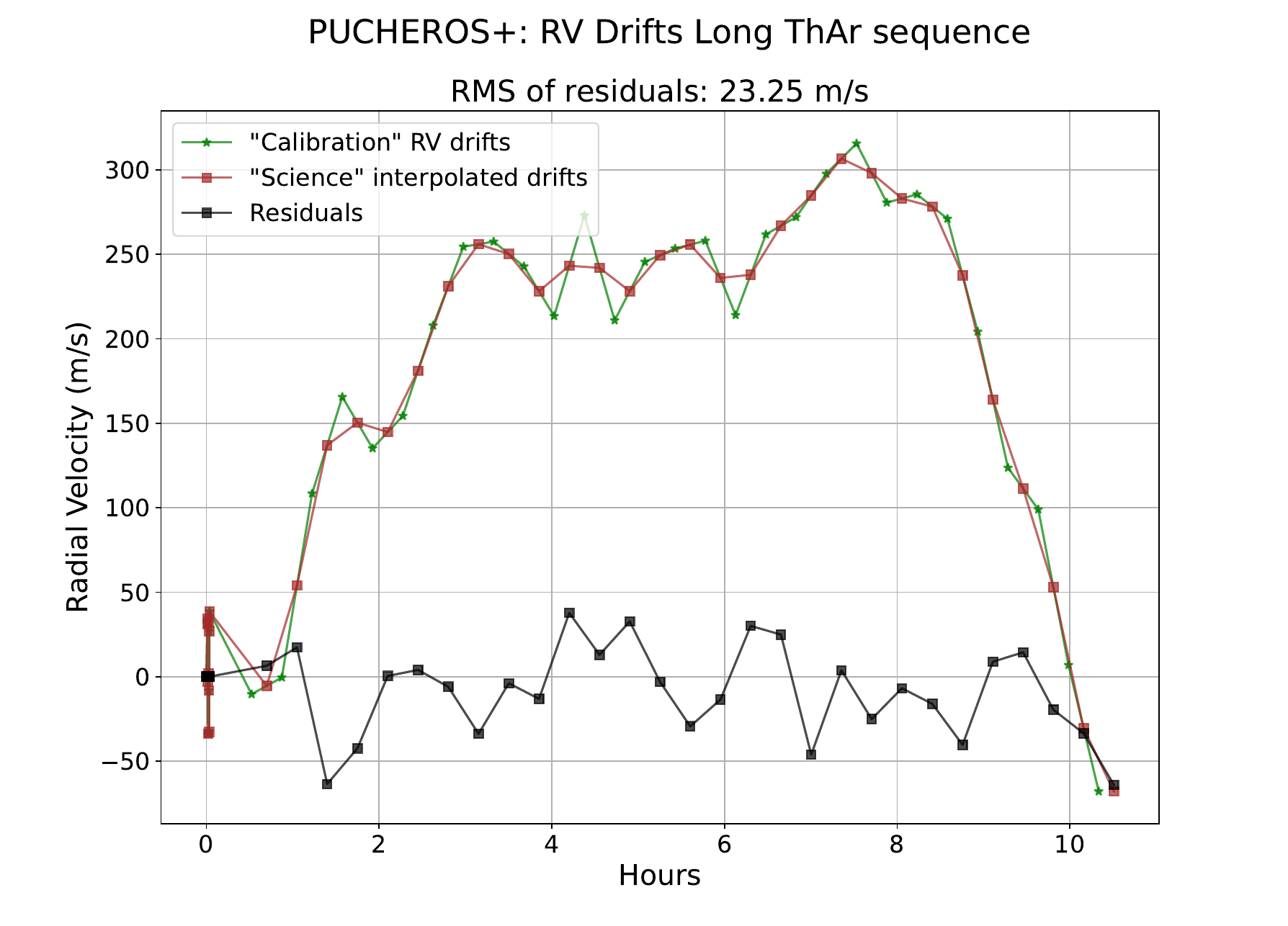}
    \caption{Simulating 10 minutes observation, intercalating ThAr spectra, a maximum precision achievable of 23.3\,m\,s$^{-1}$ is found.}
    \label{fig:MaximumPrecision1}
    \end{subfigure}
    
    \begin{subfigure}[b]{1\linewidth}
        \centering
    \includegraphics[width=1\linewidth]{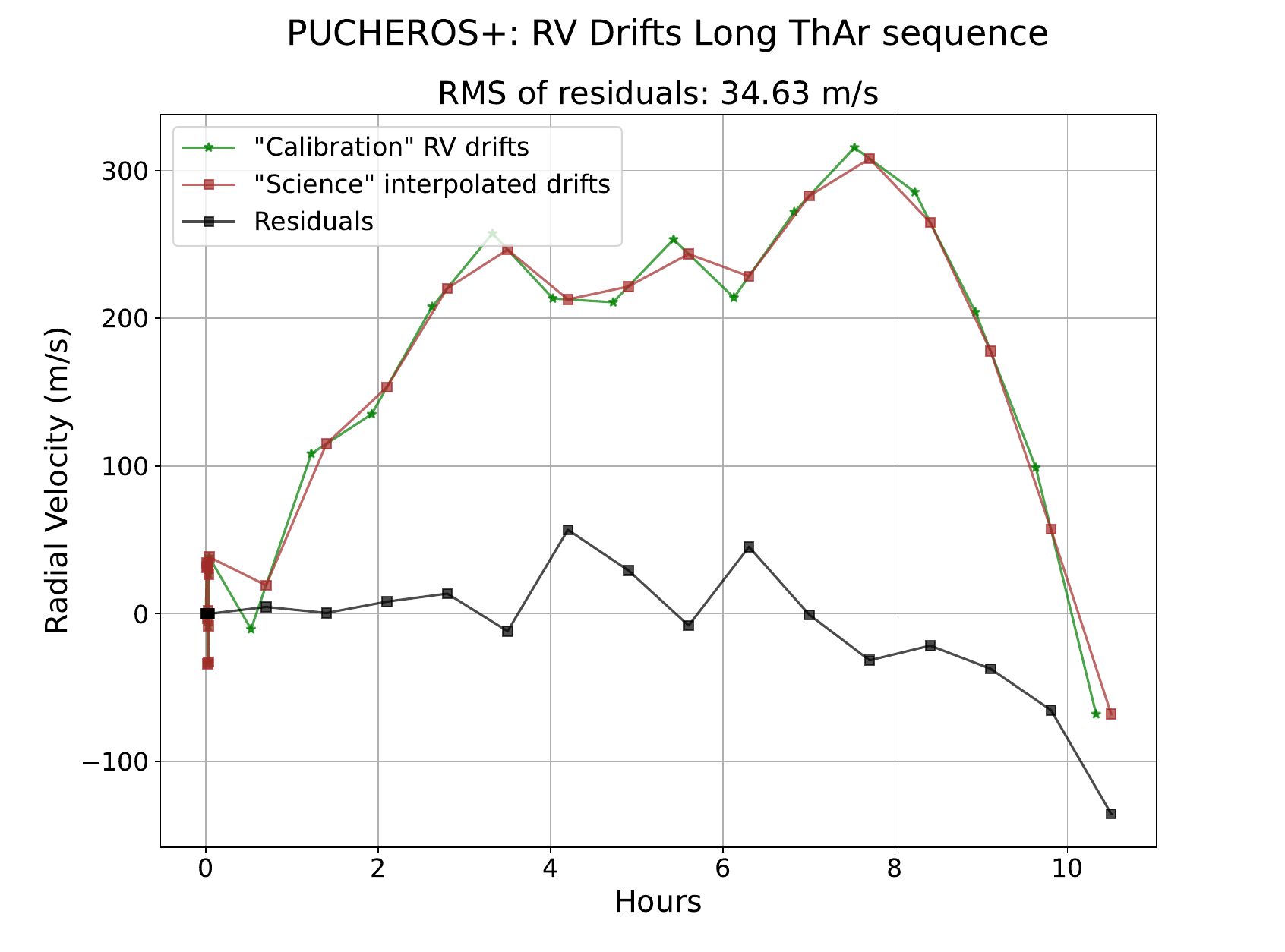}
    \caption{Simulating 30 minutes observation, intercalating ThAr spectra, a maximum precision achievable of 34.6\,m\,s$^{-1}$ is found.}
    \label{fig:MaximumPrecision2}
    \end{subfigure}

    \caption{Radial velocity of the instrumental drift of a $10$ hours ThAr sequence: green curve are the "calibration" drifts, brown curve is the linear interpolation from "calibration" drift, the black curve are the residuals.}
    \label{fig:MaximumPrecision}
\end{figure}

These results can be contrasted with real radial velocity measurements of standard stars. We analyze data from two radial velocity standard stars.
The first RV-standard, HD171990, was observed, during 4 nights, with several observations each night. In this case (Figure \ref{fig:HD171990}), the precision achieved throughout the nights was of $38$\,m\,s$^{-1}$ (considering 48 images). The results for each individual night are the following:  $39.7$\,m\,s$^{-1}$, $45.1$\,m\,s$^{-1}$, $28.8$\,m\,s$^{-1}$ and $32.8$\,m\,s$^{-1}$. 

\begin{figure}
    \centering
    \includegraphics[width=1\linewidth]{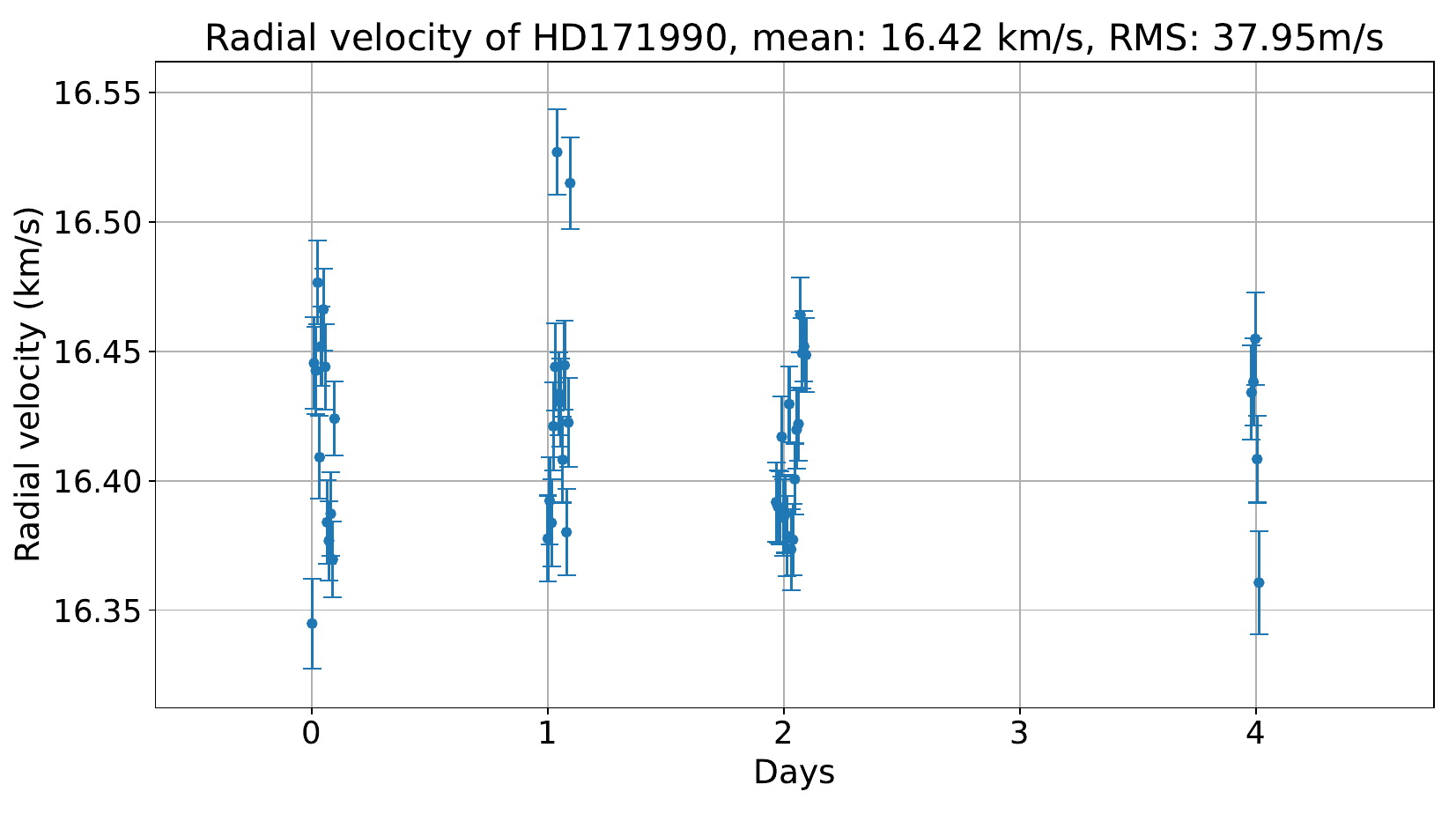}
    \caption{Radial velocity measurements for RV standard HD171990 over four different nights of observations.}
    \label{fig:HD171990}
\end{figure}

The other RV standard, HD100623, was observed 20 times over an interval of 23 nights. The precision achieved over 10 nights is $33.9$\,m\,s$^{-1}$ (Figure \ref{fig:HD100623subfig2}), consistent with the previous result, while over the full period is $55.9$\,m\,s$^{-1}$ (Figure \ref{fig:HD100623subfig1}). 
Therefore, based on the observation of RV-standard stars HD171990 and HD100623, we find a measured precision of $30\sim 40$\,m\,s$^{-1}$  up to 10 nights of observation for PUCHEROS+.
 
\begin{figure}
    \centering
    \begin{subfigure}[b]{1\linewidth}
        \centering
        \includegraphics[width=1\linewidth]{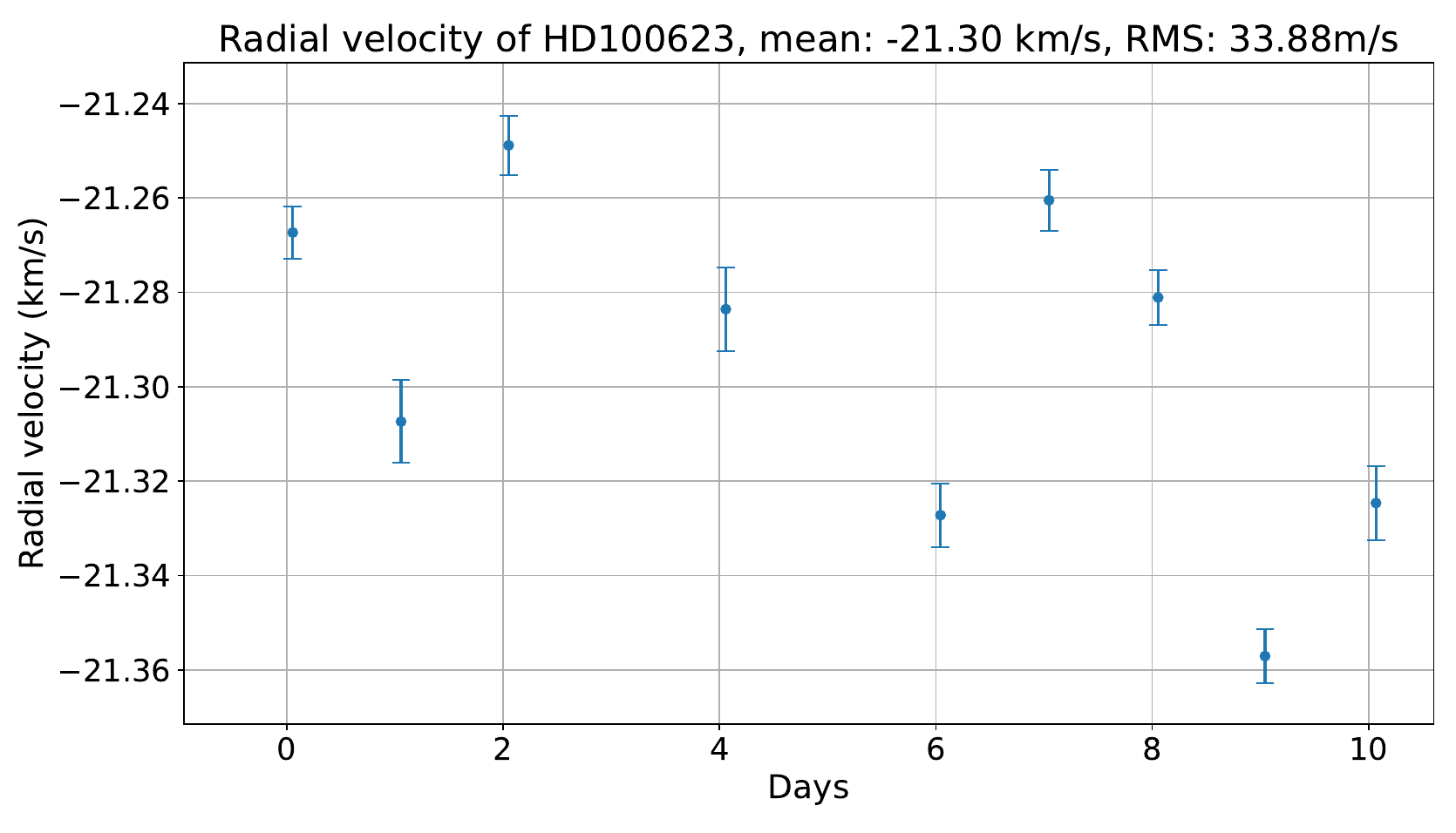}
        \caption{Over 10 nights of observations.}
        \label{fig:HD100623subfig2}
    \end{subfigure}

    \begin{subfigure}[b]{1\linewidth}
        \centering
        \includegraphics[width=1\linewidth]{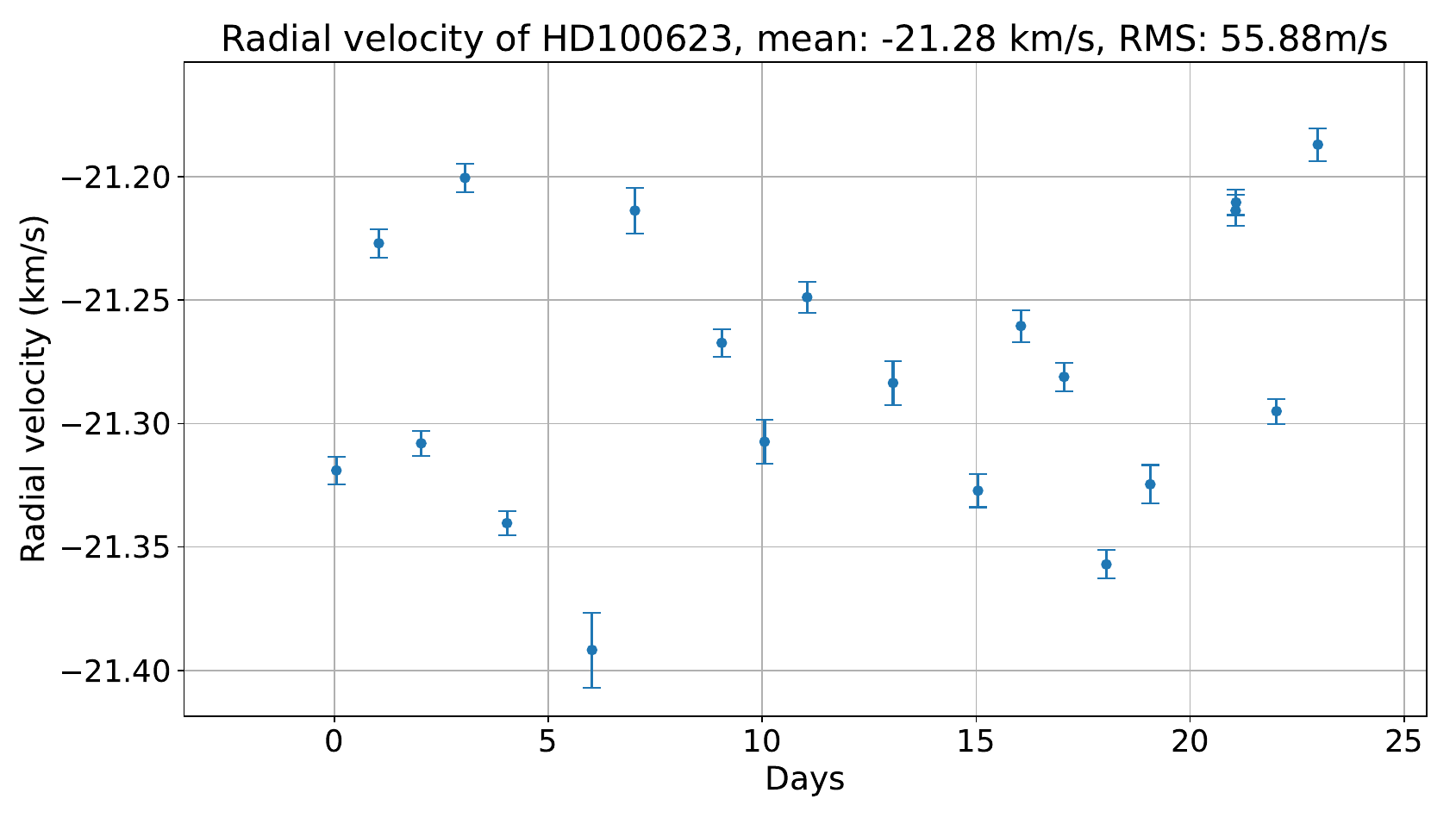}
        \caption{Over 23 nights of observations.}
        \label{fig:HD100623subfig1}
    \end{subfigure}

    \caption{Radial velocity measurements for RV standard HD100623 over several nights.}
    \label{fig:HD100623}
\end{figure}

\section{Science capabilities} \label{section: science}

In this section, five successful scientific cases are presented, highlighting the scientific capabilities of the spectrograph.

\subsection{Orbit of binary star BM Mon}

BM\,Mon is an eclipsing binary with an orbital period of about $1.245$ days \citep{skarka2017}. The visual magnitude is $12.2$. The primary component, with a temperature of more than $11{,}000\,K$, is of spectral type B. From the shape of the light curve obtained by \textit{TESS} mission \citep{Ricker2014}, we are estimating the temperature of a secondary component to be around $5{,}000\,K$ (spectral type K).

We obtained 8 usable {spectra from PUCHEROS+ on the ESO 1.52\,m telescope during October 2024. Based on the observing conditions, the values of SNR vary from $10$ to $30$ for $1{,}800\,s$ long exposure. We observed this target also using OES spectrograph \citep{Kabath2020} attached at Perek 2-meter telescope in Ond\v{r}ejov (spectral resolution is $50{,}000$). Weather conditions and atmospheric seeing in Central Europe are significantly worse compared to La Silla. The altitude of the observatory is also lower (around 500 m a.s.l.). Moreover, instrument performance is affected by different spectral resolutions, used coatings and camera sensitivities. Taking all these effects into account, we had to use double exposure time (i.e. 3600\,s) to obtain a similar value of SNR and spectrum quality as from PUCHEROS+, even while the Perek telescope has nearly double the collected area.

\begin{figure}
    \centering
    \includegraphics[width=1\linewidth]{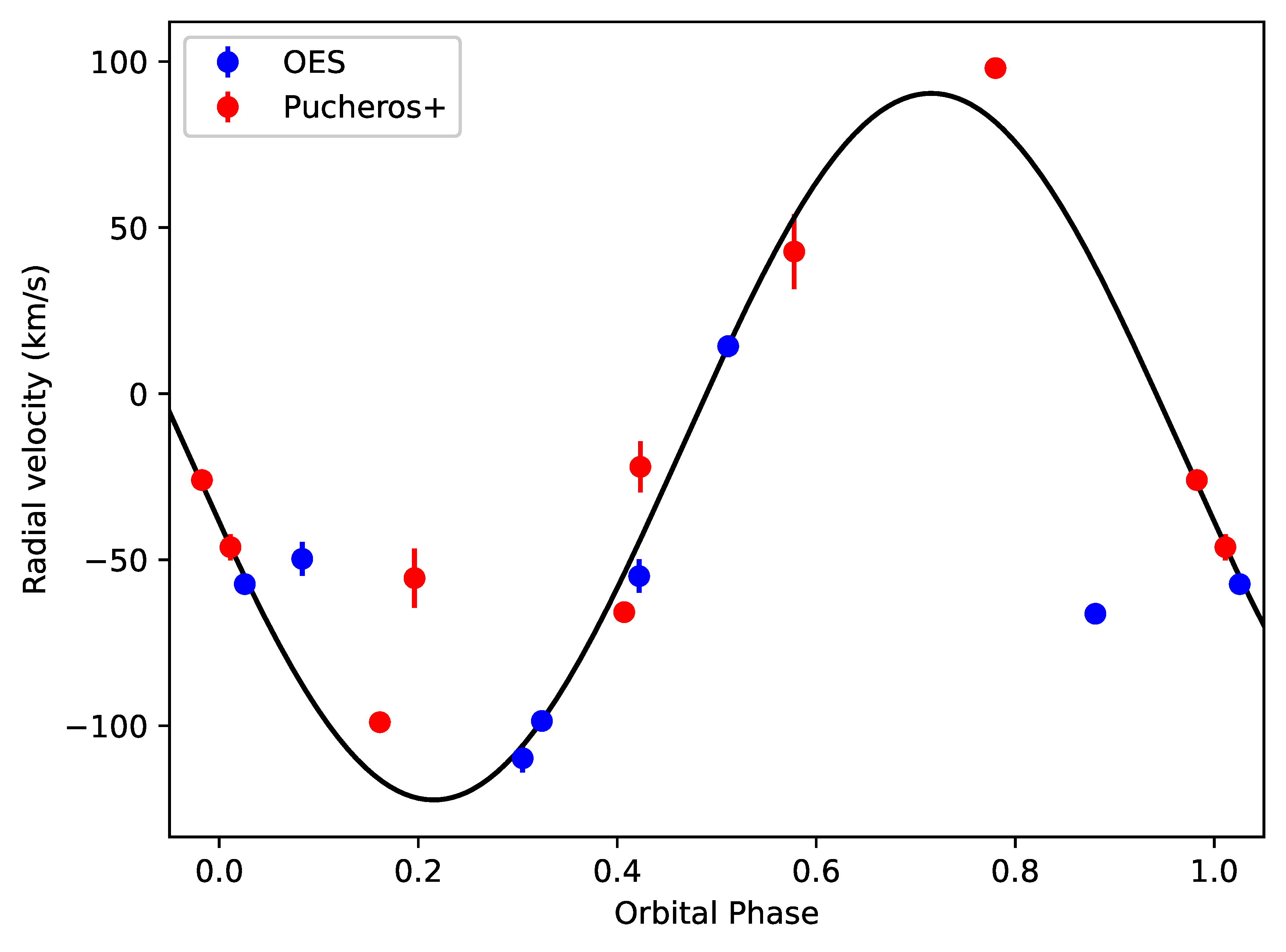}
    \caption{The radial velocities of BM Mon measured with OES and PUCHEROS+ spectrographs. Some error bars are too small to be visible.}
    \label{fig:BMMon}
\end{figure}

There were only a few lines in the spectrum of such a hot star to precisely measure the RVs. We selected the 24$^{\rm th}$ order of reduced PUCHEROS+ spectrum covering the range from 500 to 512\,nm with silicon Si II lines. Using synthetic template \citep{Coelho2014}, we were able to obtain RVs of the primary component with a mean uncertainty of about 1\,km\,s$^{-1}$

Fitting of the RV curve using a circular orbit (Figure\,\ref{fig:BMMon}) gives an amplitude of $106\pm6$\,km\,s$^{-1}$. The mass function is 0.16\,M$_\odot$ which gives a mass ratio of about 0.5 for the main sequence B-type star as a primary component. This value agrees with our rough estimations of the binary parameters based on the \textit{TESS} light curve.

This case proves the usability of the PUCHEROS+ spectrograph in studying complicated targets from the RV point of view -- a faint, hot star with limited strong lines.

\subsection{High-amplitude \texorpdfstring{$\delta$}{delta} Sct star AI Vel}

AI\,Vel is one of a well-studied prototypes of high-amplitude $\delta$\,Sct stars \citep[HADS,][]{Breger2000} that has been extensively studied by means of photometry and served as a testing laboratory for stellar pulsations and mode identifications \citep[e.g.][]{Walraven1955,Walraven1992,Guzik1993,Petersen1996}. These population I stars pulsate in radial and non-radial pressure modes. Particularly, AI\,Vel pulsates in the fundamental ($f_{\rm F}=8.9627$\,c/d) and first overtone modes ($f_{\rm 1O}=11.5997$\,c/d), and also shows signatures of non-radial modes \citep{Walraven1992}. 

We used PUCHEROS+ at the ESO 1.52\,m telescope to get radial velocities (RVs) of AI\,Vel, and gathered 64 spectra on February 15, 2024. Already \citet{Balona1980} showed that the radial velocities are almost perfectly in anti-phase with the photometric variations which goes against the expectation that there is a 0.1 cycle shift between maximum light and minimum of RVs \citep{Breger1976}. Our observations confirm no phase shift (still in anti-phase). We used light-curve model of AI\,Vel based on the \textit{TESS} satellite photometry to get photometric variations in the time of spectroscopic observations since AI\,Vel has not been monitored by \textit{TESS} in this period (see Figure\,\ref{fig:AI Vel}). As can be seen from Figure\,\ref{fig:AI Vel}, the radial velocities seem to reflect the photometric variations, though not exactly. The differences may be given either by the fact that the photometric model is based on data taken a year before the spectroscopic observations and the pulsation properties may have slightly changed during this period, or that the differences are real. 

\begin{figure}
    \centering
    \includegraphics[width=1\linewidth]{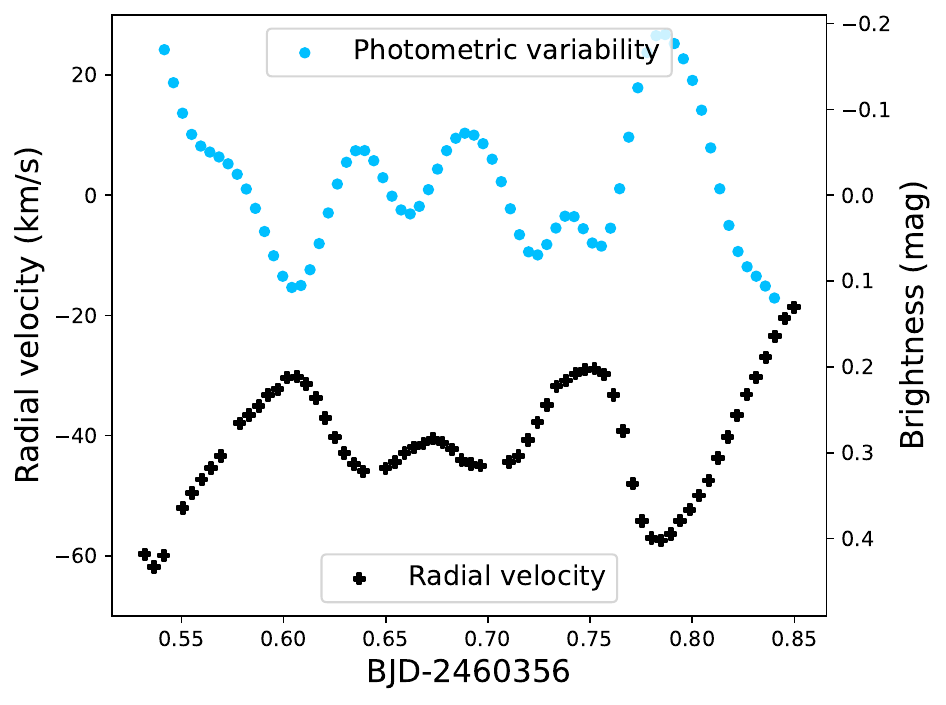}
    \caption{Radial velocity of AI Vel measured with PUCHEROS+ (black crosses) together with expected photometric variations based on the model from the \textit{TESS} satellite. The RVs are shifted of -60 km\,s$^{-1}$ for a better readability.}
    \label{fig:AI Vel}
\end{figure}

This simple demonstration shows that stability and precision of PUCHEROS+ spectra allows us to study fine details in $\delta$\,Sct stars, HADS and pulsating stars in general that have not been possible with more noisy data. An additional example of application of PUCHEROS+ observations can be a better definition of the RV to photometric amplitudes ratio that is usually between 50 and 125\,km\,s$^{-1}$\,mag$^{-1}$ \citep{Breger1976} and investigation of the phase shifts between RVs and light variations. For AI\,Vel, the RV to light-amplitude ratio is about 100\,km\,s$^{-1}$\,mag$^{-1}$ and there is no apparent phase shift between RVs minima and photometric maxima (see Figure\,\ref{fig:AI Vel}).

\subsection{Rossiter-McLaughlin effect of HAT-P-41 Ab}
The Rossiter-McLaughlin effect is a spectroscopic anomaly that manifests as anomalous radial velocities due to spectral line deformation during a transit \citep{ross24,mcl,alb22}. This effect allows us to determine the projected angle between the stellar spin axis and the normal to the orbital plane. This angle helps us to understand the migration pathways of exoplanets as aligned planets are usually linked to slow disc migration without any violent effects in the history of the system, while polar and retrograde orbits hint at violent dynamical processes such as planet-planet and star-planet interactions. These dynamical events can bring the planet initially on a distant orbit beyond the snowline much closer to the host star via the Kozai-Lidov mechanism \citep{fab07}. 

HAT-P-41 Ab is a hot Jupiter with an extended atmosphere around an F-type star on 2.7-day orbit \citep{hart12}. The host is a bright (11.1 \textit{V}mag) star and is part of a binary system where HAT-P-41 B is located at a projected separation of $1{,}240$ au. Thanks to its large planetary radius and relatively fast stellar rotation ($v\,\sin i_*=16.5 \pm 0.6$ km\,s$^{-1}$), the system represents an excellent target for measuring the Rossiter-McLaughlin effect. 
HAT-P-41 was observed with PUCHEROS+ during three transit windows in the summer of 2024. We have used exposure times between $1{,}680\,s$ and $1{,}800\,s$ obtaining 35 frames with SNR between 35 and 65 which resulted in radial velocity uncertainties between $93$ and $230$\,m\,s$^{-1}$ and a mean uncertainty of $125$\,m\,s$^{-1}$.
To measure the projected stellar obliquity of the system ($\lambda$), we follow the
methodology of \citet{zak24}: we fit the RVs with a composite model, which includes a Keplerian orbital component as
well as the R-M anomaly. This model is implemented in the
\textsc{ARoMEpy}\footnote{\url{https://github.com/esedagha/ARoMEpy}} \citep{seda23} package, which utilizes the
\textsc{Radvel} \texttt{python} module \citep{ful18} for the formulation
of the Keplerian orbit. \textsc{ARoMEpy} is a \texttt{python} implementation
of the R-M anomaly described in the \textsc{ARoME} code \citep{bou13}. We set gaussian priors on the projected rotational velocity $v\,\sin i_*$ and fit for the projected obliquity. The best fit values were obtained using the Markov chain Monte Carlo method. We derive a prograde orbit with $\lambda =4^{+25}_{-31}$\,deg. Such a result is in good agreement with already published obliquity measurement using the HARPS-N data yielding $\lambda =-4.4^{+5.0}_{-5.6}$\,deg \citep{zak25}. We show the observed radial velocity and the best fit obtained with PUCHEROS+ data in Figure\,\ref{fig:RM HAT-P-41}.
The obtained result shows the capability of PUCHEROS+ and potential for similar instruments where the flexibility in scheduling can be used to obtain many nights to improve the signal. The extended \textit{TESS} and the upcoming PLATO mission are expected to discover many more gas giants around bright stars amenable for Rossiter-McLaughlin effect characterization with small and medium size telescopes. PUCHEROS+ like instruments can be used for the initial screening of the spin-orbit angle, with the most interesting being later followed up by instruments on larger telescopes.

\begin{figure}
    \centering
    \includegraphics[width=1\linewidth]{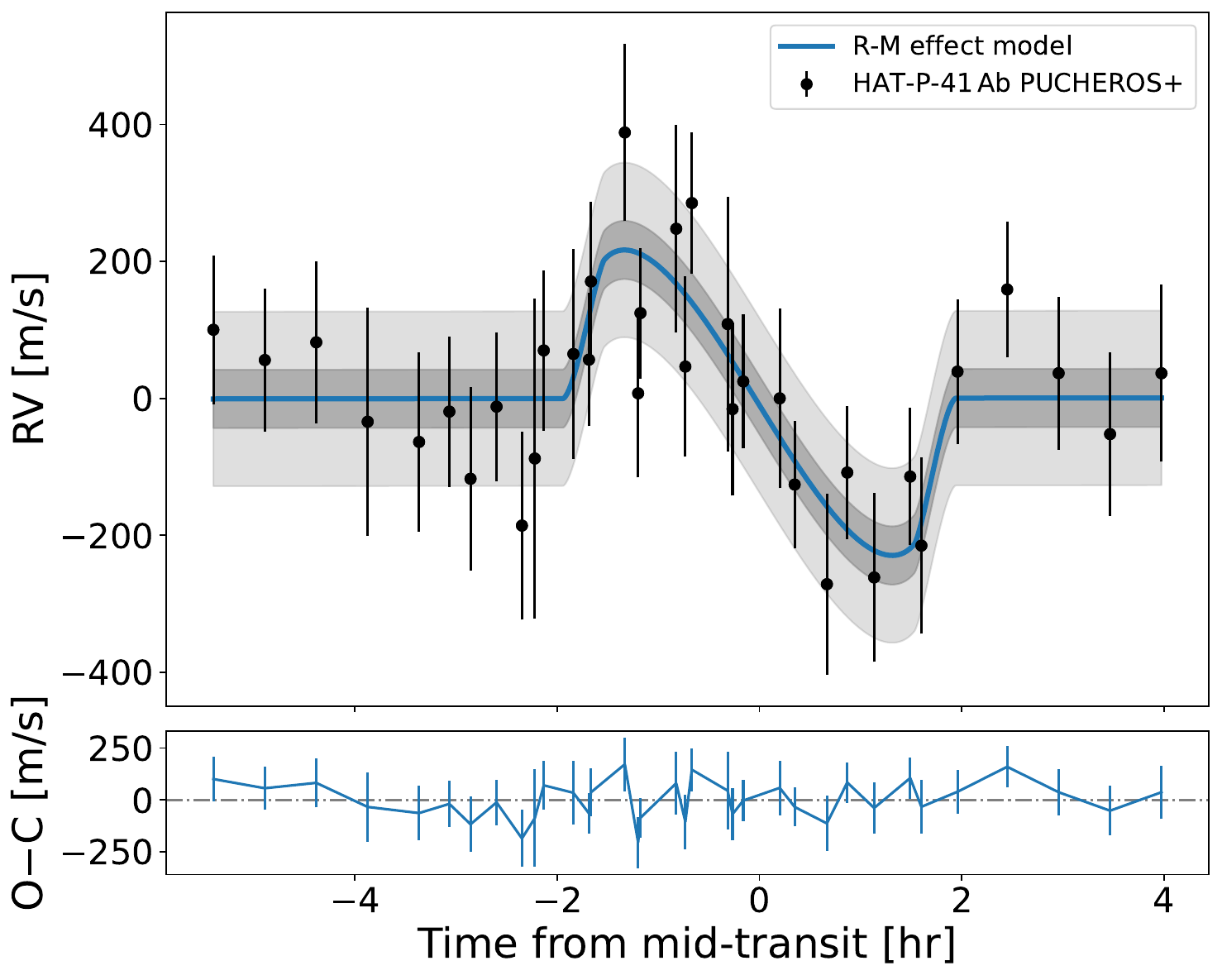}
    \caption{Rossiter-Mclaughlin effect of planet HAT-P-41Ab measured by PUCHEROS+. We derive an aligned orbit consistent with the previous literature value.}
    \label{fig:RM HAT-P-41}
\end{figure}

\subsection{Flares on EQ Peg}
\label{sec:flares}
Stellar flare research has become increasingly enhanced since the space-based photometry from \textit{Kepler} and \textit{TESS} became available, which allowed detailed statistics on this transient phenomenon on a large variety of stars. This also allowed the detection of so-called superflares with energies larger than $10^{33}$\,erg and showed this to be a rather common phenomenon, even on solar-like stars \citep{Maehara2012}. However, compared to photometric observations, spectroscopic observations of stellar flares are much rarer due to the required time investment, but these are vital to assess the physical parameters of flares. Between 2022 and 2024, we monitored a sample of flaring G, K, and M stars, selected according to their flare rates in \textit{TESS} data, with PUCHEROS+ at the ESO 1.52\,m telescope. In total, more than $8{,}000$ spectra were recorded for this campaign, and the first results focusing on the active young M dwarf AU\,Mic were published in \citet{Odert2025}.

Here we show an example of a flare observation on the active M dwarf binary EQ\,Peg. This system consists of an M3.5Ve primary and a M4.5Ve secondary separated by $\sim$6\arcsec\ \citep{Crosley2018}. Both components are well-known flare stars. An analysis of \textit{TESS} sectors\,56 and 83 revealed a flare rate of about 1.17 per day in the 2\,min cadence data and 2.38 per day for the shorter 20\,s cadence. Most of the PUCHEROS+ observations of EQ\,Peg were scheduled in September 2024 during \textit{TESS} sector 83, focusing on the brighter component A. Additional photometry in the g'-band was taken simultaneously with OndCam installed at one of the two finder telescopes at the ESO 1.52\,m.

In Figure\,\ref{fig:eqpeg}, we show one example of a flare on EQ\,Peg\,A occurring on 2024-09-24. The binary system is unresolved in the photometric observations, but since the flare is observed in the spectra of A at the same time, we can also attribute the photometric flare to A. The flare was clearly detected in several chromospheric lines, and as is commonly observed in M dwarfs, the H$\alpha$ emission has a much longer decay phase than the white-light component measured with the broad-band photometry. Whereas the flare lasts only about 0.5\,h in white-light, the duration in H$\alpha$ is almost 3\,h. The peak flux amplitude of the H$\alpha$ flare is 5\,per cent and its integrated energy about $3\times10^{30}$\,erg. In comparison, the amplitude of the white-light component is 0.8\,per cent in the \textit{TESS} band and 8\,per cent in the g'-band. From the photometric observations we estimate flare energies of ${\sim}3\times10^{31}$\,erg in the \textit{TESS} band and ${\sim}8\times10^{32}$\,erg in the g'-band, and under the assumption that the flare emission corresponds to a $\sim$ $9{,}000$\,K blackbody its total radiative output could be a few times $10^{33}$\,erg, which puts it in the superflare regime.

These results demonstrate the usage of PUCHEROS+ for the characterization of stellar flares and superflares, especially in conjunction with ground- and/or space-based photometry, or even as part of multi-wavelength observation campaigns.

\begin{figure}
    \centering
    \includegraphics[width=1\linewidth]{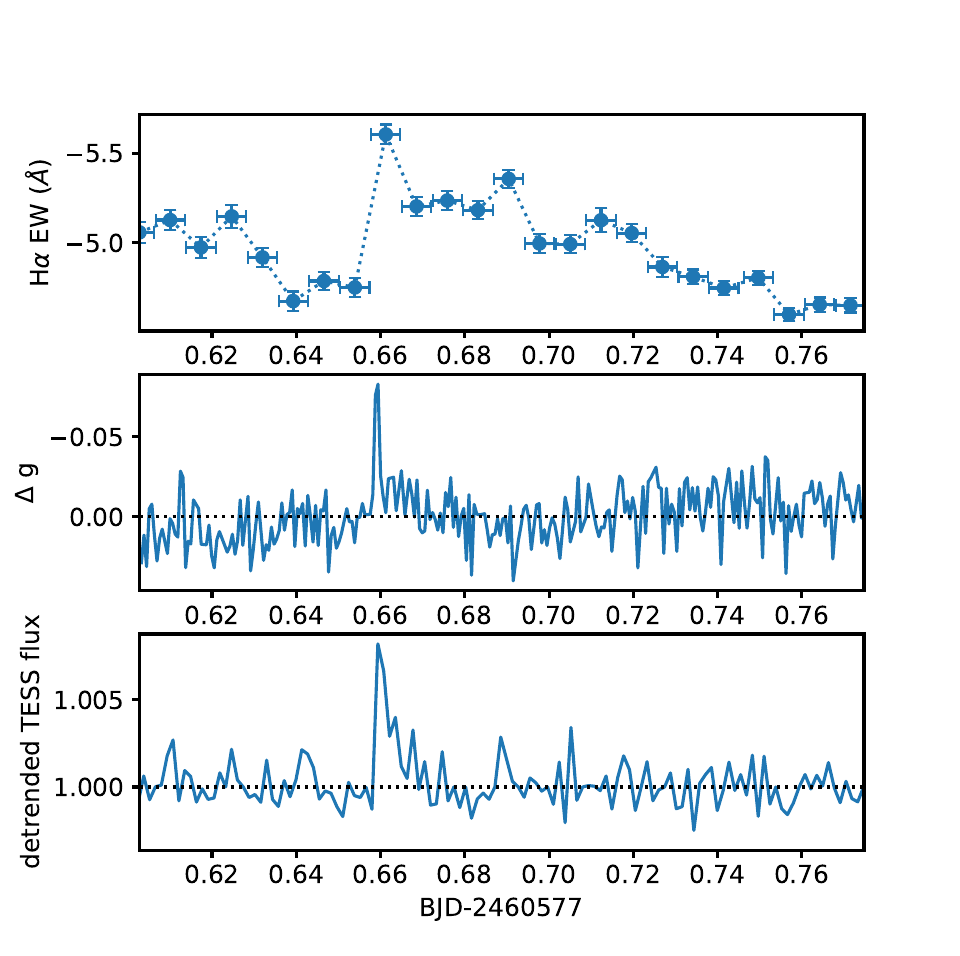}
    \caption{Flare peaking at BJD-2460577=0.66 on EQ Peg A observed simultaneously with PUCHEROS+ (upper panel: H$\alpha$ equivalent width), OndCam (middle panel: differential g'-band magnitude), and \textit{TESS} (lower panel: normalized detrended \textit{TESS} flux).}
    \label{fig:eqpeg}
\end{figure}

\subsection{Slingshot prominences on PZ Tel}
Slingshot prominences have first been detected on the fast-rotating K dwarf AB\,Dor as transient absorption features drifting from blue to red across the H$\alpha$ line \citep{Robinson1986, CollierCameron1989}. They are interpreted as cool plasma confined in magnetic field loops which are centrifugally extended in the equatorial plane and rotate rigidly with the star \citep{Jardine2001}. Later this phenomenon has been detected on other fast-rotating G, K, and M stars \citep[e.g.][]{Doyle1990,Barnes2000,Dunstone2006a}. Moreover, it was detected also in other chromospheric lines such as \ion{Ca}{ii}\,HK and in some cases, the off-disk emission of these structures could also be observed \citep{Dunstone2006b}. As the fast-rotating stars forming such prominences are generally magnetically active, several of them were included in the flaring star sample described in section\,\ref{sec:flares} and thus observed with PUCHEROS+ in the course of this campaign. We show a dynamic residual spectrum in the wavelength region of the H$\alpha$ line of the fast-rotating young G9/K0 star PZ\,Tel in Figure\,\ref{fig:pztel}, showing a prominent absorption feature between BJD-2460000=59.76-59.82 drifting from the blue to the red wing of the line. Measuring the drift rate and duration of the absorption allows to constrain the geometric parameters of the prominence \citep{Leitzinger2016}. The slope of the prominence track indicates that the structure is located around the co-rotation radius of the star. Using a cloud model we estimate the maximum prominence area of 10\% of the stellar disk assuming the structure is optically thick. These parameters are comparable to those obtained in earlier studies of PZ\,Tel \citep{Barnes2000,Leitzinger2016}. This demonstrates the capabilities of PUCHEROS+ for studying stellar slingshot prominences.

\begin{figure}
    \centering
    \includegraphics[width=1\linewidth]{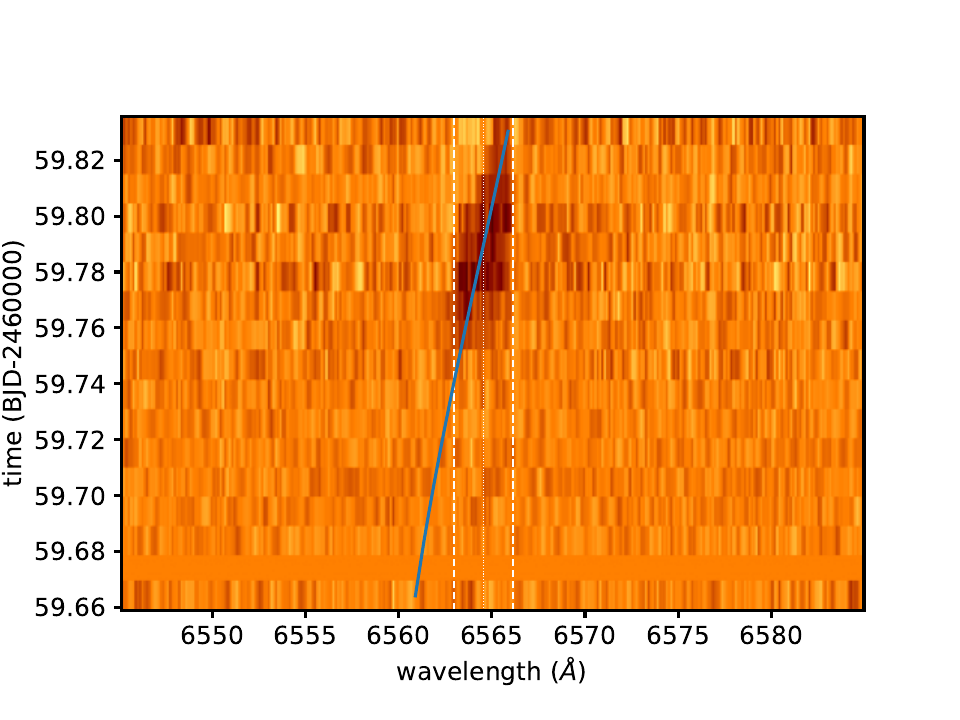}
    \caption{Dynamic residual spectrum of PZ Tel around the H$\alpha$ line. A slingshot prominence can be seen as dark absorption feature between BJD-2460000=59.76-59.82, drifting from blue to red with time. The dotted line indicates the line center and the dashed lines the $\pm v\sin{i}$ range. A prominence track is displayed as a solid line.}
    \label{fig:pztel}
\end{figure}

\section{Discussion} \label{section: sec5 Discussion}
In this section, we briefly compare PUCHEROS+ with PUCHEROS in terms of costs and performance.

Regarding the components, the detector of PUCHEROS is not available anymore, however the difference in cost with the detector of PUCHEROS+ is about a factor 4, roughly $10{,}000$ USD versus $40{,}000$ USD. For the objective, the difference is about a factor 2, the objective of PUCHEROS was made of two commercial lenses whose current cost is about $400$ USD, while the photographic objective of PUCHEROS+ is about $800$ USD.
The thermal control was absent in PUCHEROS and it represents a cost of about $2{,}500$ USD.
Including a flat continuum lamp in the telescope interface represents a minor cost. The rest of the components were kept unchanged.

In terms of the performance, the RV precision is improved by more than a factor of 3, about $30$\,m\,s$^{-1}$ in PUCHEROS+ while it was not better than $100$\,m\,s$^{-1}$ in PUCHEROS.

In terms of limiting magnitude, we compare Figure \ref{fig:ETC_frontend_old} with Figure 5 of \citep{Vanzi2012} - calculated assuming an efficiency of 4\%. For instance, PUCHEROS at the UC Santa Martina 50-cm telescope gives SNR = 50 on a magnitude V = 7 star in 300 seconds, PUCHEROS+ at the ESO 1.52\,m telescope gives a similar result on a V = 10 star. Based on simple scaling up of the telescope diameter, we would expect to gain 2.4 magnitudes, the extra gain of about 0.6 magnitudes can be attributed to the increased efficiency of the detector and the reduction of pinhole losses because of the improved quality of the observing site. 

We also place special emphasis on the system's automation and remote observing capabilities, which make PUCHEROS+ a powerful spectrograph for scientific or even public observatories.

The improvements in the spectrograph performance enable new science that the original PUCHEROS was unable to perform. In particular, the observation of fainter stars thanks to the higher efficiency and the absence of Residual
Bulk Image effect in the new detector; and the detection of giant exo-planets, which is possible thanks to the precision of $30\sim 40$ m\,s$^{-1}$; are among the most significant scientific advances made possible by PUCHEROS+.

\section{Conclusions} \label{section: sec6 Conclusions}

In this paper, we presented the design and performance of PUCHEROS+, an enhanced version of PUCHEROS spectrograph. The instrument was tested at the ESO 1.52\,m telescope of La Silla, and was operated remotely since its installation on 2022. The comparison and improvement with respect to PUCHEROS were described in detail. We found that the new instrument overcomes the original limitations, it improves the RV precision by a factor of 5 reaching about $30$m\,s$^{-1}$, and the limiting magnitude is 3 magnitudes deeper. These enhancements make PUCHEROS+ an interesting instrument for a number of scientific applications including the study of binary and multiple systems, stellar activity, astroseismology, transients targets and exoplanets. In addition, PUCHEROS+ was a successful test bench for our observing protocols implemented later for PLATOSpec.

\section*{Acknowledgements}

We thank support from the PLATOSpec consortium and support from ESO staff during the installation and operation of the instrument in La Silla. We thank logistic support from the staff of the AIUC and in particular Maria Eliana Escobar and Gishline Sade. This work was supported by ANID Fondecyt n. 1211162, Quimal ASTRO20-0025, and Basal CATA FB210003. 
We thank all the observers who contributed to collect the data from PUCHEROS+.

J.L. and J.Z. acknowledge support from GACR grant 22-30516K under which some data were obtained. 
R.B. acknowledges support from FONDECYT Project 1241963 and from ANID -- Millennium  Science  Initiative -- ICN12\_009.
Furthermore, P.K., P.G. and M.S. acknowledge support from LTT-20015 for the variable stars data sets.
The research of P.G. was supported by the Slovak Research and Development Agency under contract No. APVV-20-0148 and the internal grant No. VVGS-2023-2784 of the P. J. {\v S}af{\'a}rik University in Ko{\v s}ice funded by the EU NextGenerationEU through the Recovery and Resilience Plan for Slovakia under the project No. 09I03-03-V05-00008.
This research was funded in whole, or in part, by the Austrian Science Fund (FWF) [10.55776/I5711, 10.55776/P37256, 10.55776/PAT4657624]. This paper includes data collected by the TESS mission, which are publicly available from the Mikulski Archive for Space Telescopes (MAST). Funding for the TESS mission is provided by the NASA’s Science Mission Directorate.
\section*{Data Availability}
The data supporting this article will be shared on a reasonable request to the corresponding author.



\bibliographystyle{mnras}
\bibliography{pucherosp_references} 




%
%
%
%
%
\bsp	
\label{lastpage}
\end{document}